\documentclass[aip,preprint,notitlepage,amsmath,amssymb]{revtex4-2}
\usepackage[T1]{fontenc}
\usepackage{lmodern}
\usepackage[english]{babel}
\usepackage{graphicx}
\usepackage[caption=false,subrefformat=parens]{subfig}
\usepackage{microtype}
\usepackage{placeins}
\usepackage{setspace}
\usepackage[hidelinks]{hyperref}
\hypersetup{pdftitle={Conductivity of electrolyte solutions: Self-consistent Debye-Huckel-Onsager theory},
  pdfauthor={Yury A. Budkov and Nikolai N. Kalikin}}
\AtBeginDocument{\singlespacing\raggedbottom}
\newcommand{\angstrom}{\mbox{\normalfont\AA}}
\newcommand{\celsius}{\ensuremath{\,^{\circ}\mathrm{C}}}

\begin{document}

\author{\firstname{Yury A.} \surname{Budkov}}
\email[]{ybudkov@hse.ru}
\affiliation{Laboratory of Computational Physics, HSE University, Tallinskaya st. 34, 123458 Moscow, Russia}
\affiliation{School of Applied Mathematics, HSE University, Tallinskaya st. 34, 123458 Moscow, Russia}
\affiliation{Laboratory of Multiscale Modeling of Molecular Systems, G.A. Krestov Institute of Solution Chemistry of the Russian Academy of Sciences, 153045, Akademicheskaya st. 1, Ivanovo, Russia}
\author{\firstname{Nikolai N.} \surname{Kalikin}}
\affiliation{Laboratory of Computational Physics, HSE University, Tallinskaya st. 34, 123458 Moscow, Russia}
\affiliation{Laboratory of Multiscale Modeling of Molecular Systems, G.A. Krestov Institute of Solution Chemistry of the Russian Academy of Sciences, 153045, Akademicheskaya st. 1, Ivanovo, Russia}

\title{Conductivity of electrolyte solutions: Self-consistent Debye--H\"uckel--Onsager theory}
\date{20 November 2026}

\begin{abstract}
We develop a self-consistent Debye--H\"uckel--Onsager (SCDHO) theory for the electrical conductivity of electrolyte solutions. Ion-specific Slater form factors describe spatially distributed charges and regularize the Coulomb interaction at short distances. We obtain the steady-state distortion of the ionic atmosphere from the gradients of local chemical potentials in classical density functional theory and the Ornstein--Zernike equation. The conductivity separates into mean-field, correlation (relaxation), and electrophoretic contributions. In the random phase approximation, the two interaction corrections reduce to one-dimensional integrals. We fit charge smearing lengths to measured molar conductivities for aqueous electrolytes of several valence types, using a common length for each ion across the fitted solutions. These lengths are then used to calculate the temperature dependence without further fitting. Comparisons with measurements in several non-aqueous solvents illustrate how the model extends to different polar media.
\end{abstract}

\maketitle

\noindent Journal reference: \href{https://doi.org/10.1039/D5CP04521E}{\textit{Phys.\ Chem.\ Chem.\ Phys.}\ \textbf{28}, 6756--6768 (2026)}.

\section{Introduction}
Electrolyte solutions play a fundamental role in a wide range of natural and technological systems. They are essential for vital biological processes, such as nerve signalling and muscle contraction~\cite{schoen2007mechanism,bohr1964electrolytes}, as well as for maintaining fluid balance in the body~\cite{metheny2012fluid}. In technology, they serve as the basis for energy storage in batteries, and they are critical components of desalination and water purification processes that address the global need for fresh water~\cite{alkhadra2022electrochemical,chaplin2019prospect}. For these applications, highly concentrated electrolyte solutions play a crucial role, but the theoretical basis for their behavior is currently underdeveloped. Accurately predicting the electrical conductivity of these solutions is a significant challenge for physicists, chemists, and engineers, with applications in various fields such as biology, medicine, energetics, and chemical engineering.

The theoretical basis for understanding electrolyte conductivity was laid down in the pioneering works of Debye, H\"uckel, Onsager and others~\cite{onsager1927report,debye1923theory,onsager1927theory,onsager2002irreversible,fuoss1978review,barthel1968conductance}. Recently, this field has seen a revival driven by advances in statistical mechanics and nonequilibrium thermodynamics, which have led to the development of phenomenological nonequilibrium models~\cite{fraenkel2010simplified,fraenkel2018improved,kimura2024revisit,boroujeni2023estimation} and the application of sophisticated statistical-mechanical methods such as integral equations for correlation functions~\cite{bernard1992conductance,bernard2023analytical,naseri2023new}, stochastic density functional theory (SDFT)~\cite{demery2016conductivity,avni2022conductance,avni2022conductivity_,berthoumieux2024nonlinear,illien2025dean}, and mean-field theory~\cite{vinogradova2023electrophoresis,vinogradova2025conductivity}. These advanced theories have been successful in simulating conductivity over a wide range of concentrations.

Despite these significant advances, a critical aspect of ion electrostatics remains largely unaddressed in the context of conductivity: the non-local nature of the ionic charge distribution. From a quantum-mechanical perspective, the charge of a solvated ion is neither localized at a point nor confined to a thin spherical shell, but is instead distributed over a finite volume, resembling a diffuse cloud. This non-locality is expected to become significant at high concentrations where the mean interionic distance is small enough for the point-charge approximation to break down. While recent theoretical studies on out-of-equilibrium electrolytes have effectively incorporated short-range modifications of the Coulomb potential~\cite{avni2022conductance,avni2022conductivity_,robin2024correlation}, these empirical corrections have not been systematically linked to the underlying physical origin -- the quantum-mechanically non-local charge distribution of solvated ions. Deriving such form factors from first principles remains a formidable challenge. However, valuable insights can be drawn from several fields. In plasma physics~\cite{kraeft1986quantum}, quantum effects introduce a renormalization cut-off at short distances. Similarly, phenomenological models and microscopic statistical field theories provide frameworks for describing non-local electrostatic effects~\cite{wang2010fluctuation,budkov2020statistical,budkov2024statistical,kornyshev1996shape,rubashkin2015nonlocal}. Within the latter approach, recent empirical charge form factors for ions have been proposed to describe equilibrium properties of ionic fluids~\cite{frydel2013double,frydel2016double,adar2019screening}.

A widely used model for charge smearing in coarse-grained simulations of soft matter, as described in the literature~\cite{mao2015modeling,warren2014screening,ibergay2009electrostatic,chiacchiera2024coarse}, is the Slater-type form factor. We recently applied this charge smearing model to the description of the electrical conductivity of aqueous electrolyte solutions within the framework of the modified Debye--H\"uckel--Onsager theory~\cite{kalikin2024modified}. Although successful, this approach of using the same Slater-type charge smearing function for both cation and anion is fundamentally unsatisfactory. Different ions have distinct electronic structures and solvation environments, requiring a model that assigns unique form factors and smearing lengths for each solvated ion. Thus, the charge smearing length should serve as {a solvent-dependent parameter} for each ion in a polar solvent, capturing its unique chemical properties. Additionally, we would like to reduce our reliance on semi-empirical data while maintaining the model's predictive power.

In this work, we address these gaps by developing a new self-consistent Debye--H\"uckel--Onsager (SCDHO) theory that explicitly incorporates the non-local charge distribution of ions through a Slater-type charge form factor, leading to a regularized Coulomb potential at short ranges. In contrast to the classical Debye--H\"uckel--Onsager theory and its subsequent modifications, this paper uses the expressions for the probability flows associated with the gradient of the local chemical potentials of ionic species. These expressions are based on the principles of classical density functional theory and the Ornstein--Zernike equation for the correlation functions. We use this framework to derive analytical expressions for the electrophoretic and correlation (relaxation) contributions to the total electric conductivity, enabling us to model the electrical conductivity of various aqueous and non-aqueous electrolyte solutions across a wide range of ion valences, concentrations, and temperatures.

\section{Self-consistent Debye--H\"uckel--Onsager theory}

\subsection{Basic equations}

{In the spirit of the Onsager--Fuoss treatment~\cite{onsager2002irreversible}, we consider a sufficiently large and homogeneous system containing, in the most general case, $s$ species of ions, indexed by $a=1,\dots,s$ with mean number densities $\bar{c}_a$. The ions carry total charges $q_a$. The solvent is treated as a continuous, incompressible fluid with shear viscosity $\eta$ and dielectric permittivity $\varepsilon$. An external, constant, homogeneous electric field $\boldsymbol{E}$ is applied, driving the ions into a steady state of conduction.}

Following Onsager and Fuoss~\cite{onsager2002irreversible,fuoss1978review}, we adopt a two-particle correlation approximation for the continuity equation within the configuration space of ions in an electrolyte solution. Namely, in the configuration space of two ions, $a$ and $b$, in a steady state, which is the sole focus of this work, the following continuity equation for the probability flows holds~\cite{pitaevskii2012physical}
\begin{equation}
\label{continuous}
\nabla_a\cdot \boldsymbol{j}_{ab}(\boldsymbol{r}_a,\boldsymbol{r}_b) +\nabla_{b}\cdot\boldsymbol{j}_{ba}(\boldsymbol{r}_b,\boldsymbol{r}_a)=0,
\end{equation}
where the operators $\nabla_a$ and $\nabla_b$ mean that differentiation is performed over the ion positions $\boldsymbol{r}_a$ and $\boldsymbol{r}_b$, respectively;
\begin{equation}
\label{flux}
\boldsymbol{j}_{ab}(\boldsymbol{r}_a,\boldsymbol{r}_b)=g_{ab}(\boldsymbol{r}_a,\boldsymbol{r}_b)\mathbf{v}_{ab}(\boldsymbol{r}_a,\boldsymbol{r}_b),\quad a\leftrightarrow b
\end{equation}
is the conditional probability flow for ion $a$ at $\boldsymbol{r}_a$ when ion $b$ is fixed at $\boldsymbol{r}_b$. The function $g_{ab}(\boldsymbol{r}_a,\boldsymbol{r}_b)$ is the steady-state pair distribution function. The total velocity of ions $a$ around ion $b$ is
\begin{equation}
\label{velocity}
\boldsymbol{v}_{ab}(\boldsymbol{r}_a,\boldsymbol{r}_b)=m_a q_a\boldsymbol{E}-m_a\nabla_a \mu_{a}(\boldsymbol{r}_a,\boldsymbol{r}_b)+\boldsymbol{u}_b(\boldsymbol{r}_a,\boldsymbol{r}_b),\quad a\leftrightarrow b,
\end{equation}
where $m_a$ is the mobility of ion $a$, $q_a$ is its total electric charge, $\mu_{a}(\boldsymbol{r}_a,\boldsymbol{r}_b)$ is the local chemical potential of ion $a$ which is a functional of the local concentrations of ionic species, $c_a(\boldsymbol{r}_a,\boldsymbol{r}_b)=\bar{c}_ag_{ab}(\boldsymbol{r}_a,\boldsymbol{r}_b)$; $\boldsymbol{u}_b(\boldsymbol{r}_a,\boldsymbol{r}_b)$ is the average solvent velocity field at point $\boldsymbol{r}_a$ around ion $b$ placed at point $\boldsymbol{r}_b$.
The total nonequilibrium chemical potential of ions $a$ can be written as the sum of an ideal-gas term and an excess term as follows\footnote{It is important to note that the chemical potential lacks a rigorous definition in non-equilibrium states. However, for systems close to equilibrium, the concept can be extended by treating the chemical potential as a functional of non-equilibrium concentrations. See also our recent paper~\cite{budkov2025fluctuation}.}:
\begin{equation}
\label{chemical_pot_full}
\mu_{a}(\boldsymbol{r}_a,\boldsymbol{r}_b)=k_B T\ln(\lambda_a^3\bar{c}_ag_{ab}(\boldsymbol{r}_a,\boldsymbol{r}_b))+\mu^{(\text{ex})}_{a}(\boldsymbol{r}_a,\boldsymbol{r}_b),\quad a\leftrightarrow b,
\end{equation}
Here $k_B$ is the Boltzmann constant, $T$ is the temperature, and $\lambda_a$ is the thermal de Broglie wavelength of species $a$.

Further, assuming that the ion $b$ at point $\boldsymbol{r}_b$ causes a rather small perturbation of the solution environment, we obtain to first order in the concentration perturbations
\begin{equation}
\label{chemical_pot_linear}
\mu_{a}(\boldsymbol{r}_a,\boldsymbol{r}_b)\approx \bar{\mu}_{a}+k_BTh_{ab}(\boldsymbol{r}_a,\boldsymbol{r}_b)-k_{B}T \sum\limits_{l}\int d^3\boldsymbol{r}'c_{al}(\boldsymbol{r}_a-\boldsymbol{r}')\delta c_{l}(\boldsymbol{r}',\boldsymbol{r}_b),
\end{equation}
where
\begin{equation}
\bar{\mu}_{a}=\bar{\mu}^{(\text{id})}_{a}+\bar{\mu}^{(\text{ex})}_{a}
\end{equation}
is the bulk chemical potential of ions $a$ with $\bar{\mu}^{(\text{id})}_{a}=k_B T\ln(\lambda_a^3\bar{c}_a)$ and $\bar{\mu}^{(\text{ex})}_{a}$ being the ideal-gas contribution and excess contribution, respectively,
\begin{equation}
h_{ab}(\boldsymbol{r}_a,\boldsymbol{r}_b)=g_{ab}(\boldsymbol{r}_a,\boldsymbol{r}_b)-1
\end{equation}
is the steady-state total correlation function of ions $a$ and $b$,
and
\begin{equation}
c_{ab}(\boldsymbol{r}-\boldsymbol{r}')=-\frac{\delta (\beta\mu^{(\text{ex})}_{a}(\boldsymbol{r},\boldsymbol{r}_b))}{\delta c_{b}(\boldsymbol{r}',\boldsymbol{r}_b)}\bigg{|}_{\text{bulk}}
\end{equation}
is the direct correlation function evaluated in the homogeneous bulk solution, with $\beta=(k_B T)^{-1}$.
The concentration perturbations of ions $a$ at point $\boldsymbol{r}$ caused by the occurrence of ion $b$ at point $\boldsymbol{r}'$ are
\begin{equation}
\label{delta_c}
\delta c_{a}(\boldsymbol{r},\boldsymbol{r}')=
\begin{cases}
\bar{c}_{a}h_{ab}(\boldsymbol{r},\boldsymbol{r}'), & \text{if}\quad  a\neq b\,\\
\bar{c}_{b}h_{bb}(\boldsymbol{r},\boldsymbol{r}')+\delta(\boldsymbol{r}-\boldsymbol{r}'),& \text{if} \quad a= b.
\end{cases}
\end{equation}
It is important to mention that the total correlation functions satisfy the following symmetry relation:
\begin{equation}
\label{trans}
h_{ab}(\boldsymbol{r}_a,\boldsymbol{r}_b)=h_{ba}(\boldsymbol{r}_b,\boldsymbol{r}_a).
\end{equation}

Thereby, substituting Eq.~\eqref{delta_c} into Eq.~\eqref{chemical_pot_linear}, we can obtain the chemical potential in the linear approximation with respect to $h_{ab}$ as follows
\begin{equation}
\label{chemical_pot_2}
\mu_{a}(\boldsymbol{r}_a,\boldsymbol{r}_b)= \bar{\mu}_{a}+k_BTh_{ab}(\boldsymbol{r}_a,\boldsymbol{r}_b)-k_B T c_{ab}(\boldsymbol{r}_a-\boldsymbol{r}_b)-k_{B}T\int d^3\boldsymbol{r}' \sum\limits_{l}c_{al}(\boldsymbol{r}_a-\boldsymbol{r}') \bar{c}_{l} h_{lb}(\boldsymbol{r}',\boldsymbol{r}_b).
\end{equation}
Further, representing the total correlation function as a sum of equilibrium, $\bar{h}_{ab}$, and steady-state, $\omega_{ab}$, parts
\begin{equation}
h_{ab}(\boldsymbol{r},\boldsymbol{r}')=\bar{h}_{ab}(\boldsymbol{r}-\boldsymbol{r}')+\omega_{ab}(\boldsymbol{r},\boldsymbol{r}'),
\end{equation}
taking into account the Ornstein--Zernike (OZ) equation~\cite{barrat2003basic,budkov2025fluctuation}
\begin{equation}
\label{OZ}
\bar{h}_{ab}(\boldsymbol{r}_a-\boldsymbol{r}_b)= c_{ab}(\boldsymbol{r}_a-\boldsymbol{r}_b)+\int d^3\boldsymbol{r}'\sum\limits_{l}c_{al}(\boldsymbol{r}_a-\boldsymbol{r}') \bar{c}_{l} \bar{h}_{lb}(\boldsymbol{r}'-\boldsymbol{r}_b),
\end{equation}
and substituting (\ref{chemical_pot_2}) into (\ref{velocity}), and subsequently  (\ref{velocity}) into the continuity equation (\ref{continuous}), we obtain the following linear basic equations for the steady-state part of the total correlation functions
\begin{multline}
\label{basic}
m_a k_B T \nabla_a^2 \hat{\omega}_{ab}(\boldsymbol{r}_a,\boldsymbol{r}_b)+m_b k_B T \nabla_b^2 \hat{\omega}_{ba}(\boldsymbol{r}_b,\boldsymbol{r}_a)\\=(\boldsymbol{u}_b(\boldsymbol{r}_a,\boldsymbol{r}_b)+q_a m_a\boldsymbol{E})\cdot\nabla_a{h}_{ab}(\boldsymbol{r}_a,\boldsymbol{r}_b)+(\boldsymbol{u}_a(\boldsymbol{r}_b,\boldsymbol{r}_a)+q_b m_b\boldsymbol{E})\cdot\nabla_b{h}_{ba}(\boldsymbol{r}_b,\boldsymbol{r}_a),
\end{multline}
where we have introduced the following shorthand
\begin{equation}
\hat{\omega}_{ab}(\boldsymbol{r}_a,\boldsymbol{r}_b)=\omega_{ab}(\boldsymbol{r}_a,\boldsymbol{r}_b)-\int d^3\boldsymbol{r}'\sum\limits_{l}c_{al}(\boldsymbol{r}_a-\boldsymbol{r}') \bar{c}_{l} \omega_{lb}(\boldsymbol{r}',\boldsymbol{r}_b),
\end{equation}
took into account that in the linear approximation in $h_{ab}$ in the continuity equation (\ref{continuous}) $g_{ab}(\boldsymbol{r}_a,\boldsymbol{r}_b)\approx 1$, and used the condition for the velocity field
\begin{equation}
\label{incompres}
\nabla_a \cdot\boldsymbol{u}_b(\boldsymbol{r}_a,\boldsymbol{r}_b)=\nabla_b \cdot\boldsymbol{u}_a(\boldsymbol{r}_b,\boldsymbol{r}_a)=0,
\end{equation}
which follows from the incompressibility of the solvent.

In order to close eqs. (\ref{basic}), it is necessary to formulate a hydrodynamic equation for the velocity field. For steady state and relatively weak solvent flow, we can use the linearized Navier--Stokes equation (the Stokes equation)~\cite{pitaevskii2012physical,vinogradova2023electrophoresis,vinogradova2025conductivity,avni2022conductance,avni2022conductivity_}
\begin{equation}
\label{Stokes}
\eta \nabla_a^2 \boldsymbol{u}_b(\boldsymbol{r}_a,\boldsymbol{r}_b)-\nabla_a p_b(\boldsymbol{r}_a,\boldsymbol{r}_b)+\boldsymbol{E} \varrho_b(\boldsymbol{r}_a,\boldsymbol{r}_b)=0,\quad a \leftrightarrow b
\end{equation}
where $p_b(\boldsymbol{r}_a, \boldsymbol{r}_b)$ is the pressure of the solvent at point $\boldsymbol{r}_a$, near ion $b$, which is placed at point $\boldsymbol{r}_b$,
\begin{equation}
\varrho_b(\boldsymbol{r}_a,\boldsymbol{r}_b)=\sum\limits_{l}\bar{c}_l\hat{q}_l h_{lb}(\boldsymbol{r}_a,\boldsymbol{r}_b),\quad a \leftrightarrow b
\end{equation}
is the local electric charge density of ions at point $\boldsymbol{r}_a$, near ion $b$ (and {\sl vice versa}), where we have introduced the charge smearing integral operator (see discussion below)~\cite{wang2010fluctuation,frydel2013double,frydel2016double}
\begin{equation}
\hat{q}_l h_{lb}(\boldsymbol{r}_a,\boldsymbol{r}_b)=\int d^3\boldsymbol{r}' \hat{q}_{l}(\boldsymbol{r}_a-\boldsymbol{r}')h_{lb}(\boldsymbol{r}',\boldsymbol{r}_b).
\end{equation}
Its kernel, $\hat{q}_{a}(\boldsymbol{r})$, represents the charge density of ion $a$, which satisfies the normalization condition $\int d^3\boldsymbol{r}\hat{q}_{a}(\boldsymbol{r}) =q_a$.

{To first order in the total correlation functions $h_{ab}$ on the right-hand side of Eq.~\eqref{basic}, we can safely omit the solvent velocity field $\boldsymbol{u}_b$ since it is itself linear in the total correlation function and thus produces the quadratic terms in $h_{ab}$ in the basic equations (\ref{basic}). Physically, this means that we ignore the influence of electrophoretic flow on the steady-state correlation functions, as was done in classical DHO theory~\cite{fuoss1978review,pitaevskii2012physical}.} In this approximation the correlation functions are translation invariant, that is $\omega_{ab}(\boldsymbol{r}_a,\boldsymbol{r}_b)=\omega_{ab}(\boldsymbol{r}_a-\boldsymbol{r}_b)$. In the linear approximation on the electric field, we can approximate the total correlation function on the right-hand side of (\ref{basic}) as follows ${h}_{ab}(\boldsymbol{r}_a,\boldsymbol{r}_b)=\bar{h}_{ab}(\boldsymbol{r}_a-\boldsymbol{r}_b)$. Thus, taking into account that $\boldsymbol{r}_a-\boldsymbol{r}_b=\boldsymbol{r}$ and that $\nabla_a=-\nabla_b=\nabla$ we can rewrite the basic equations (\ref{basic}) in the Fourier representation as follows
\begin{multline}
\label{basic_}
(m_a+m_b)k_B Tk^2{\omega}_{ab}(\boldsymbol{k}) -m_a k_B T k^2 \sum\limits_{l} c_{al}(\boldsymbol{k})\bar{c}_l\omega_{lb}(\boldsymbol{k})\\-m_b k_B T k^2 \sum\limits_{l} c_{bl}(\boldsymbol{k})\bar{c}_l\omega_{la}(-\boldsymbol{k})=-(q_a m_a-q_b m_b)i(\boldsymbol{E}\cdot \boldsymbol{k}) \bar{h}_{ab}(\boldsymbol{k}),
\end{multline}
where we have introduced the Fourier transforms of the functions in accordance with the standard rule $f(\boldsymbol{k})=\int d^3\boldsymbol{r}e^{-i\boldsymbol{k}\cdot\boldsymbol{r}}f(\boldsymbol{r})$.

In this approximation, all the other functions are translation invariant: $\boldsymbol{u}_b(\boldsymbol{r}_a,\boldsymbol{r}_b)=\boldsymbol{u}_b(\boldsymbol{r}_a-\boldsymbol{r}_b)$, $p_b(\boldsymbol{r}_a,\boldsymbol{r}_b)=p_b(\boldsymbol{r}_a-\boldsymbol{r}_b)$, $\varrho_b(\boldsymbol{r}_a,\boldsymbol{r}_b)= \bar{\varrho}_b (\boldsymbol{r}_a-\boldsymbol{r}_b)=\sum_{l}\bar{c}_l\hat{q}_l \bar{h}_{lb}(\boldsymbol{r}_a-\boldsymbol{r}_b)$. Thus, by excluding the pressure, using the solvent incompressibility condition (\ref{incompres}) after some algebra, we arrive at~\cite{pitaevskii2012physical}
\begin{equation}
\boldsymbol{u}_b(\boldsymbol{k})=\frac{1}{\eta k^2}\left(\boldsymbol{E}-\frac{\boldsymbol{k}(\boldsymbol{k}\cdot\boldsymbol{E})}{k^2}\right)\bar{\varrho}_b(\boldsymbol{k}),
\end{equation}
where we have introduced the Fourier transform of the local charge density:
\begin{equation}
\label{rho_b}
\bar{\varrho}_b(\boldsymbol{k})=\sum\limits_{l}\bar{c}_l \hat{q}_{l}(\boldsymbol{k})\bar{h}_{lb}(\boldsymbol{k}).
\end{equation}
Note that in the Stokes equation (\ref{Stokes}), we have neglected the internal force density~\cite{schmidt2022power,budkov2025fluctuation}, which in first order in $h_{ab}$ renormalizes solvent pressure and can therefore be excluded. Higher order terms in $h_{ab}$ in the internal force density can also be neglected, as their contribution exceeds the linear accuracy we have adopted.

Within the adopted linear approximation, the steady-state perturbations, $\omega_{ab}(\boldsymbol{r})$, are the scalar values with the structure $(\boldsymbol{E}\cdot \boldsymbol{r}) f(r)$, where $f(r)$ depends only on the modulus of $\boldsymbol{r}$. Therefore, $\omega_{ab}(\boldsymbol{r})$ are odd functions. Bearing in mind conditions (\ref{trans}), we arrive at the following property
\begin{equation}
\omega_{ab}(\boldsymbol{r})=\omega_{ba}(-\boldsymbol{r}),
\end{equation}
which yields the following antisymmetry condition
\begin{equation}
\label{antisymmetry}
\omega_{ab}(\boldsymbol{r})=-\omega_{ba}(\boldsymbol{r}).
\end{equation}
The latter results in $\omega_{aa}(\mathbf{r}) = 0$. This, in turn, means that the correlation function perturbations due to the application of an external field occur only for pairs of different ions.

{In the context of linear response theory, when applied to a weak and homogeneous external electric field, the total correlation function can be rigorously separated into an equilibrium component, $\bar{h}_{ab}(\mathbf{r})$, and a steady-state correction, $\omega_{ab}(\mathbf{r})$, caused by the field. This decomposition is justified physically because the mean ionic concentrations $\bar{c}_a$ remain spatially uniform under the influence of the weak and homogeneous field. The field only slightly perturbs the pair correlations, leaving the translational symmetry of the average concentrations unchanged. As a result, the equilibrium component $\bar{h}_{ab}$ corresponds to a homogeneous and isotropic reference state, and thus obeys the standard OZ equation. Any anisotropic perturbations caused by the field are fully described by the functions $\omega_{ab}$, ensuring a consistent and mathematically sound treatment within the framework of the linear response regime.}

Thus, taking into account Eq.~\eqref{antisymmetry}, we obtain the average velocity field of ionic species $a$ as follows
\begin{multline}
\label{velocity_}
\boldsymbol{v}_a= \lim\limits_{\boldsymbol{r}'\to \boldsymbol{r}} \boldsymbol{v}_{aa}(\boldsymbol{r},\boldsymbol{r}')=m_a q_{a}\boldsymbol{E}+ \lim\limits_{\boldsymbol{r}'\to \boldsymbol{r}}(\boldsymbol{u}_a(\boldsymbol{r}-\boldsymbol{r}')- m_a k_{B}T \nabla \hat{\omega}_{aa}(\boldsymbol{r}-\boldsymbol{r}'))\\= m_a q_{a}\boldsymbol{E}+m_a k_{B}T \int\frac{d^3\boldsymbol{k}}{(2\pi)^3} i\boldsymbol{k}\sum\limits_{l(\neq a)}c_{al}(\boldsymbol{k})\bar{c}_l{\omega}_{la}(\boldsymbol{k}) + \int\frac{d^3\boldsymbol{k}}{(2\pi)^3}\boldsymbol{u}_a(\boldsymbol{k}).
\end{multline}
The total average current density is
\begin{equation}
\boldsymbol{J}=\sum\limits_{a}q_a \bar{c}_a \boldsymbol{v}_a=\boldsymbol{J}^{(\text{MF})}+\boldsymbol{J}^{(\text{cor})}+\boldsymbol{J}^{(\text{ph})},
\end{equation}
where
\begin{equation}
\boldsymbol{J}^{(\text{MF})}=\sum\limits_{a}m_aq_a^2\bar{c}_a \boldsymbol{E}
\end{equation}
is the current density within the mean-field approximation,
\begin{equation}
\label{Jcor}
\boldsymbol{J}^{(\text{cor})}=k_{B}T \int\frac{d^3\boldsymbol{k}}{(2\pi)^3} i\boldsymbol{k}\sum\limits_{a,b(\neq a)}q_a m_a \bar{c}_a \bar{c}_b c_{ab}(\boldsymbol{k}){\omega}_{ba}(\boldsymbol{k})
\end{equation}
is the correlation contribution to the total current density, and
\begin{equation}
\boldsymbol{J}^{(\text{ph})}=\frac{1}{\eta}\int\frac{d^3\boldsymbol{k}}{(2\pi)^3}\frac{\sum\limits_{a}q_{a}\bar{c}_a\bar{\varrho}_a(\boldsymbol{k})}{k^2}\left(\boldsymbol{E}-\frac{\boldsymbol{k}(\boldsymbol{k}\cdot\boldsymbol{E})}{k^2}\right)
\end{equation}
is the electrophoretic one.

To calculate the correlation contribution to the current density, it is necessary to solve the basic equations (\ref{basic}) and then substitute their solution, $\omega_{ab}(\boldsymbol{k})$, into Eq.~\eqref{Jcor}. To calculate the electrophoretic current density, we only need to know the Fourier transforms of the equilibrium total correlation functions, $\bar{h}_{ab}(\boldsymbol{k})$.

The total electric conductivity, $\sigma$, is determined by
\begin{equation}
\label{sigma}
\boldsymbol{J}=\sigma \boldsymbol{E}.
\end{equation}

\subsection{One-component electrolyte solutions}
\label{IIB}

Now, let us apply the general theory we formulated above to a situation where only one type of electrolyte is present in a solution. In this case, as it follows from eqs. (\ref{basic_}) and (\ref{antisymmetry}), only one steady-state correlation function is non-zero (see also Appendix~\ref{Derivation}):
\begin{equation}
\label{omega12}\omega_{12}(\boldsymbol{k})=-\frac{(q_1m_1-q_2 m_2)i(\boldsymbol{E}\cdot \boldsymbol{k})c_{12}(\boldsymbol{k})}{k_{B}T \Delta(\boldsymbol{k})k^2(m_1 (1-\bar{c}_1 c_{11}(\boldsymbol{k}))+m_2 (1-\bar{c}_2 c_{22}(\boldsymbol{k})))},
\end{equation}
where we took into account the relation
\begin{equation}
\label{hbar12}\bar{h}_{12}(\boldsymbol{k})=\frac{c_{12}(\boldsymbol{k})}{\Delta(\boldsymbol{k})},
\end{equation}
which follows from the OZ equation, written in the Fourier representation
\begin{equation}
\label{OZ_}
\bar{h}_{ab}(\boldsymbol{k})=c_{ab}(\boldsymbol{k}) + \sum\limits_{l}c_{al}(\boldsymbol{k})\bar{c}_l\bar{h}_{lb}(\boldsymbol{k}),
\end{equation}
and introduced the following shorthand
\begin{equation}
\Delta(\boldsymbol{k})=1-\bar{c}_1 c_{11}(\boldsymbol{k})-\bar{c}_2 c_{22}(\boldsymbol{k})+\bar{c}_1\bar{c}_2(c_{11}(\boldsymbol{k})c_{22}(\boldsymbol{k})-c_{12}(\boldsymbol{k})^2).
\end{equation}

The correlation contribution to the current density is
\begin{equation}
\boldsymbol{J}^{(\text{cor})}=-\frac{\bar{c}_1\bar{c}_2(q_1m_1-q_2m_2)^2}{3} \boldsymbol{E}\int\frac{d^3\boldsymbol{k}}{(2\pi)^3} \frac{c_{12}(\boldsymbol{k})^2}{\Delta(\boldsymbol{k})(m_1 (1-\bar{c}_1 c_{11}(\boldsymbol{k}))+m_2 (1-\bar{c}_2 c_{22}(\boldsymbol{k})))},
\end{equation}
where we have taken into account that $c_{ab}(\boldsymbol{k})$ depends only on the magnitude of $\boldsymbol{k}$.

The electrophoretic contribution is
\begin{equation}
\label{J_ph}
\boldsymbol{J}^{(\text{ph})}=\frac{2\boldsymbol{E}}{3\eta}\int\frac{d^3\boldsymbol{k}}{(2\pi)^3}\frac{\sum\limits_{a}q_{a}\bar{c}_a\bar{\varrho}_a(\boldsymbol{k})}{k^2},
\end{equation}
where we have also taken into account that $\bar{\varrho}_{a}(\boldsymbol{k})$ depends only on the magnitude of $\boldsymbol{k}$.

Therefore, the total electric conductivity, as it follows from (\ref{sigma}), is
\begin{equation}
\label{sigma_tot}
\sigma =\sigma_{\text{MF}}+\Delta\sigma_{\text{cor}}+\Delta\sigma_{\text{ph}},
\end{equation}
where the mean-field (Stokes) contribution to the total electric conductivity is
\begin{equation}
\label{sigma_mf}
\sigma_{\text{MF}}=\sum\limits_{a}m_aq_a^2\bar{c}_a,
\end{equation}
the correlation contribution is
\begin{equation}
\label{sigma_corr}
\Delta\sigma_{\text{cor}}=-\frac{(q_1m_1-q_2m_2)^2\bar{c}_1\bar{c}_2}{3}\int\frac{d^3\boldsymbol{k}}{(2\pi)^3} \frac{c_{12}(\boldsymbol{k})^2}{(m_1 (1-\bar{c}_1 c_{11}(\boldsymbol{k}))+m_2 (1-\bar{c}_2 c_{22}(\boldsymbol{k})))\Delta(\boldsymbol{k})}
\end{equation}
and the electrophoretic contribution is
\begin{equation}
\label{sigma_ph}
\Delta\sigma_{\text{ph}}=\frac{2}{3\eta}\int\frac{d^3\boldsymbol{k}}{(2\pi)^3}\frac{q_{1}\bar{c}_1\bar{\varrho}_1(\boldsymbol{k})+q_{2}\bar{c}_2\bar{\varrho}_2(\boldsymbol{k})}{k^2}.
\end{equation}
Eqs. (\ref{sigma_tot} - \ref{sigma_ph}) being the main general result of this paper solve the problem of the electric conductivity of a one-component electrolyte solution within the SCDHO theory. To calculate the conductivity, we need to know the Fourier transforms of the direct correlation functions, $c_{ab}(\boldsymbol{k})$. These can be obtained from external sources, such as integral equation theory or molecular simulations.

Note that in the random phase approximation (RPA)~\cite{budkov2024statistical}, where $c_{ab}(\boldsymbol{k})=-{\beta \tilde{V}_{ab}(\boldsymbol{k})}$  with $\beta=(k_B T)^{-1}$ and $\tilde{V}_{ab}(\boldsymbol{k})$ being the Fourier transforms of the total potentials of interactions of ions, Eq.~\eqref{sigma_corr}  becomes the one derived for by D{\'e}mery and Dean in the framework of SDFT~\cite{demery2016conductivity}:
\begin{multline}
\label{sigma_corr_Demery}
\Delta\sigma_{\text{cor}}=-\frac{(q_1m_1-q_2m_2)^2\bar{c}_1\bar{c}_2}{3(m_1+m_2)(k_B T)^2}\\
\times\int\frac{d^3\boldsymbol{k}}{(2\pi)^3}\,
\frac{\tilde{V}_{12}(\boldsymbol{k})^2}
{1+\dfrac{\beta[\bar{c}_1m_1\tilde{V}_{11}(\boldsymbol{k})+\bar{c}_2m_2\tilde{V}_{22}(\boldsymbol{k})]}{m_1+m_2}}\\
\times\bigl\{1+\beta\bar{c}_1\tilde{V}_{11}(\boldsymbol{k})+\beta\bar{c}_2\tilde{V}_{22}(\boldsymbol{k})\\
+\beta^2\bar{c}_1\bar{c}_2[\tilde{V}_{11}(\boldsymbol{k})\tilde{V}_{22}(\boldsymbol{k})-\tilde{V}_{12}(\boldsymbol{k})^2]\bigr\}^{-1}.
\end{multline}

\section{Model of smeared charges}

In the following, we will use a model of ions with smeared charges. This means that we will consider ions whose charge is distributed over a certain volume according to certain charge densities.  Specifically, the charge density at position $\boldsymbol{r}$ is given by $\hat{q}_a(\boldsymbol{r})=q_a \rho_a(\boldsymbol{r})$, where $\rho_a(\boldsymbol{r})$ is the distribution function and depends only on the distance from the origin $r=|\boldsymbol{r}|$. In this paper, we will focus on the Slater-type charge distributions, $\rho_a(r) = (8\pi l_a^3)^{-1}\exp(-r/l_a)$, with characteristic lengths $l_a$. The real-space distribution satisfies $\int d^3\boldsymbol{r}\,\rho_a(r)=1$. In the RPA,
\begin{equation}
c_{ab}(\boldsymbol{k})=-{\beta\tilde{V}_{ab}(\boldsymbol{k})}=-\frac{\beta\hat{q}_a(\boldsymbol{k})\hat{q}_b(\boldsymbol{k})}{\varepsilon k^2},
\end{equation}
where $\tilde{V}_{ab}(\boldsymbol{k})={\hat{q}_a(\boldsymbol{k})\hat{q}_b(\boldsymbol{k})}/{\varepsilon k^2}$ is the Fourier transform of the renormalized Coulomb potential of interactions of ions $a$ and $b$, $\varepsilon$ is the permittivity of solvent; $\hat{q}_a(\boldsymbol{k})=q_a \rho_a(\boldsymbol{k})$ is the Fourier transform of the charge distribution function (charge form factor) of ion $a$, where $\rho_a(\boldsymbol{k})=(1+k^2l_a^2)^{-2}$ is the Fourier transform of the Slater-type distribution function. Solving the OZ equation (\ref{OZ}), written in the Fourier representation, we obtain~\cite{kjellander1994dressed,budkov2024statistical}
\begin{equation}
\label{h_ab}
\bar{h}_{ab}(\boldsymbol{k})=-\frac{\hat{q}_a(\boldsymbol{k})\hat{q}_b(\boldsymbol{k})}{k_B T\varepsilon (k^2+\varkappa^2(\boldsymbol{k}))},
\end{equation}
where we have introduced the following screening function~\cite{budkov2024statistical,budkov2020statistical}
\begin{equation}
\varkappa^2(\boldsymbol{k})=\frac{1}{\varepsilon k_{B}T}\sum\limits_{a}\bar{c}_a\hat{q}_a(\boldsymbol{k})^2.
\end{equation}

For one-component electrolyte solutions for which $q_1=z_{1}e$ and $q_2=-z_2e$ ($z_a>0$ is the valence of species $a$, $e$ is the elementary charge) within this approximation we obtain the following correlation contribution to the conductivity
\begin{equation}
\label{sigma_corr_smeared_charge}
\Delta\sigma_{\text{cor}}=-\frac{e^6(z_1 z_2)^2(z_1m_1+z_2m_2)^2\bar{c}_1\bar{c}_2}{6\pi^2(\varepsilon k_B T)^2(m_1+m_2)}\int\limits_{0}^{\infty} \frac{dk k^2\rho_{1}(k)^2\rho_{2}(k)^2}{(k^2+\varkappa^2(k))(k^2+\varkappa_1^2(k))},
\end{equation}
where the screening function takes the form
\begin{equation}
\varkappa^2(k)=\frac{e^2}{\varepsilon k_{B}T}\left(z_1^2\bar{c}_1\rho_{1}(k)^2+z_2^2\bar{c}_2\rho_{2}(k)^2\right),
\end{equation}
and we have introduced the following auxiliary function
\begin{equation}
\varkappa_1^2(k)=\frac{e^2}{\varepsilon k_{B}T}\frac{m_1z_1^2\bar{c}_1\rho_{1}(k)^2+m_2 z_2^2\bar{c}_2\rho_{2}(k)^2}{m_1+m_2}.
\end{equation}
We have also assumed that $\hat{q}_{1}(k)=z_1 e \rho_1(k)$, $\hat{q}_{2}(k)=-z_2 e \rho_2(k)$ for the charge form-factors.

The electrophoretic contribution to the total conductivity, as it follows from eqs. (\ref{rho_b}), (\ref{sigma_ph}), and (\ref{h_ab}), is
\begin{equation}
\label{sigma_ph_smeared_charge}
\Delta\sigma_{\text{ph}}=-\frac{2 e^2}{3\pi^2\eta}\int\limits_{0}^{\infty}dk\mathcal{I}(k)\frac{\varkappa^2(k)}{k^2+\varkappa^2(k)},
\end{equation}
where $\mathcal{I}(k)=\sum\limits_{a}z_a^2\bar{c}_a\rho_a(k)/2$. Using the approximation $\mathcal{I}(k)\approx I=\sum\limits_{a}z_a^2\bar{c}_a/2$, where $I$ is the ionic strength, which is motivated by the decay of the screening factor in the integrand as $k$ increases, we obtain the simplified expression
\begin{equation}
\Delta\sigma_{\text{ph}}=-\frac{2 e^2I}{3\pi^2\eta}\int\limits_{0}^{\infty}dk\frac{\varkappa^2(k)}{k^2+\varkappa^2(k)}.
\end{equation}

\section{Numerical results}
Using eqs. (\ref{sigma_tot}), (\ref{sigma_mf}), (\ref{sigma_corr_smeared_charge}) and (\ref{sigma_ph_smeared_charge}), we now apply the developed model to reproduce experimental conductivity data for electrolyte solutions. The physical properties of the pure solvents required for the model are listed in Table~\ref{pure_solvents_properties}. Conductivity data for aqueous electrolyte solutions were primarily taken from the extensive compilation \cite{Holze2016Electrochemistry}, while the data for non-aqueous systems were sourced from \cite{janz2012nonaqueous}. The mobility, $m_a$, of the $a$-th individual ion is determined from its hydrodynamic (Stokes) radius, $R_a$, via the relation $m_a=(6\pi\eta R_a)^{-1}$. The values of the hydrodynamic radii at $25\celsius$ for aqueous systems were taken from Ref.~\cite{nightingale1959phenomenological}, whereas the influence of the temperature on the ion's radius was taken into account using Walden's rule, relying on the single ionic conductivities at infinite dilution obtained from \cite{Holze2016Electrochemistry}. For non-aqueous systems, the conductivity at infinite dilution was taken from \cite{janz2012nonaqueous} and references therein.

\begin{table}
    \centering
    \begin{tabular}{|c|c|c|}
        solvent             &   $\eta$ ($\mathrm{mPa\,s}$) \cite{Wohlfarth2016Viscosity}   &    $\varepsilon/\varepsilon_0$ \cite{Lechner2008Static}\\
        \hline
        water               &   0.89    &   78.3    \\
        dimethyl sulfoxide         &   2.01    &   47.1    \\
        N-methylformamide   &   1.75    &   176.5   \\
        methanol         &   0.55    &   32.6    \\
        ethanol   &   1.10    &   25.1   \\
    \end{tabular}
    \caption{Shear viscosity and relative permittivity of pure solvents at $25\celsius$.}
    \label{pure_solvents_properties}
\end{table}

\subsection{Aqueous electrolyte solutions}\label{aq_el_sol}
\begin{figure}[!htbp]
\centering
\subfloat[]{\includegraphics[width=0.46\linewidth]{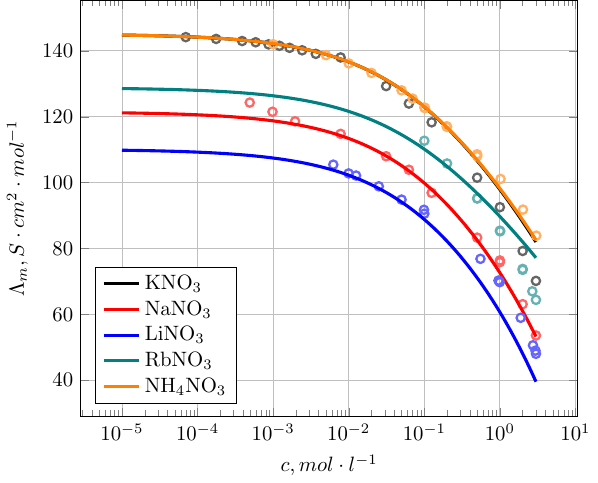}\label{1_1_nitrates}}\hfill
\subfloat[]{\includegraphics[width=0.46\linewidth]{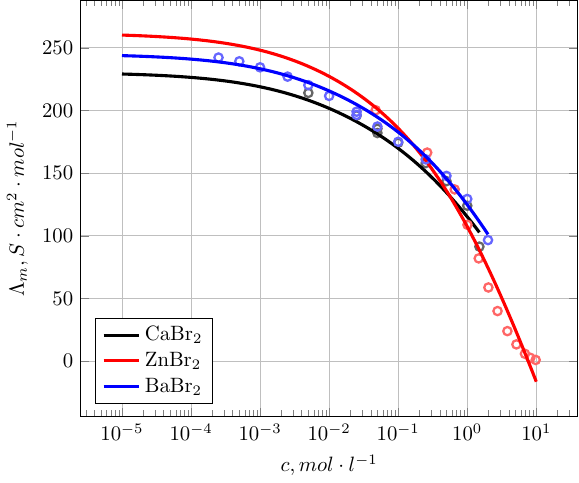}\label{2_1_bromides}}\par\smallskip
\subfloat[]{\includegraphics[width=0.46\linewidth]{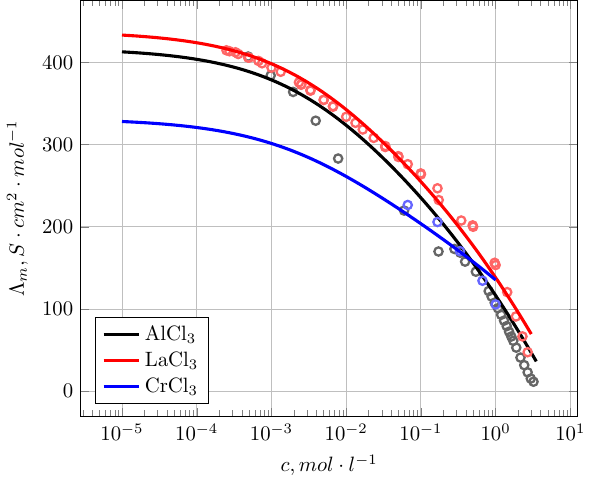}\label{3_1_chlorides}}\hfill
\subfloat[]{\includegraphics[width=0.46\linewidth]{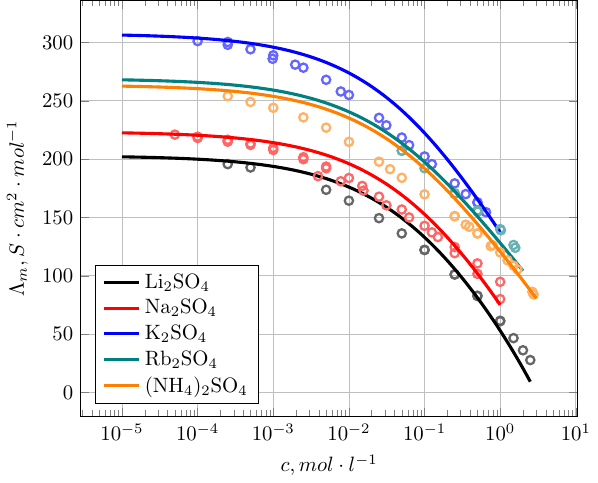}\label{1_2_sulfates}}
\caption{Molar electrical conductivity as a function of concentration for a set of aqueous electrolyte solutions: 1:1 nitrates (a), 2:1 bromides (b), 3:1 chlorides (c) and 1:2 sulfates (d).}
\label{aqueous_electrolytes}
\end{figure}

Thus, using the available dataset on the conductivity of aqueous electrolyte solutions, we simultaneously fitted the charge smearing parameters for a number of 1:1, 2:1, 3:1 and 1:2 electrolyte systems. This procedure yielded a single fitted value for each ion, which is transferable across all fitted systems. The conductivity fits for several representative systems -- including 1:1 nitrates, 2:1 bromides, 3:1 chlorides and 1:2 sulfates -- are shown in Fig.~\ref{aqueous_electrolytes} (we plot the molar conductivity, whose numerical value in $\mathrm{S\,cm^2\,mol^{-1}}$ is $10\sigma/c$ when $c$ is in $\mathrm{mol\,L^{-1}}$ and $\sigma$ is in $\mathrm{S\,m^{-1}}$); additional comparisons are given in Appendix~\ref{app:supplement}. The fitted charge smearing lengths $l_a$ for all examined ions are listed in Table~\ref{fitting_parameters}.

Overall, the shared ion-specific lengths reproduce the molar conductivity for electrolytes of different valences over a wide concentration range -- typically up to several mol/l, with a record of nearly 14 mol/l achieved for LiCl (see Appendix~\ref{app:supplement}). It is worth noting that although most data correspond to $25\celsius$, several systems were fitted at $20\celsius$ or $18\celsius$, depending on experimental data availability. However, as discussed further in Section~\ref{temp_dep_of_cond}, the charge smearing length can be considered temperature-independent over a wide temperature range.

\begin{table}
    \centering
    \begin{tabular}{c|c|c|c}
         Ion &  $l_a$, [\angstrom] & Ion & $l_a$, [\angstrom]\\
         \hline
         Ag$^+$     &   0.328    & Cr$^{3+}$       &   2.449  \\
         K$^+$      &   0.605    & Fe$^{3+}$       &   1.349  \\
          Li$^+$     &   0.302                         & Br$^-$          &   4.564 \\
         Na$^+$     &   0.385                         &  Cl$^-$          &   2.972    \\
         NH$_4$$^+$ &   0.630 &  F$^-$           &   1.146 \\
         Rb$^+$     &   0.699   & I$^-$           &   5.616  \\
         Ca$^{2+}$  &   1.116    & NO$_3$$^-$      &   1.418   \\
         Mg$^{2+}$  &   1.106    & SO$_4$$^{2-}$   &   1.223 \\
          Zn$^{2+}$  &   0.811    & CO$_3$$^{2-}$   &   1.338  \\
          Ba$^{2+}$  &   1.003    & SeO$_4$$^{2-}$  &   0.532  \\
          La$^{3+}$  &   1.494   & CrO$_4$$^{2-}$  &   1.280 \\
          Al$^{3+}$       &   1.476  &     &   \\

    \end{tabular}
    \caption{Fitted charge smearing lengths for ions in aqueous solutions.}
    \label{fitting_parameters}
\end{table}

\FloatBarrier
\subsection{Contributions to the conductivity}\label{contrib_to_conductivity}
\begin{figure}
    \centering
    \includegraphics[width=0.5\linewidth]{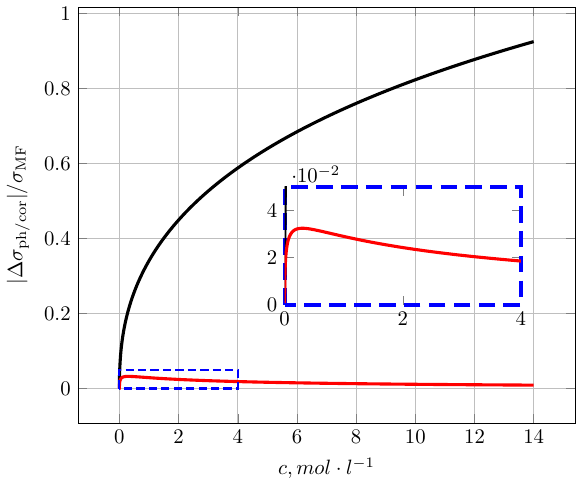}
\caption{Magnitudes of the electrophoretic (black) and correlation (red) conductivity corrections, each normalized by the mean-field conductivity, for aqueous LiCl at $25\celsius$. The inset expands the dilute region.}
    \label{LiCl_conductivity_contribution}
\end{figure}

Figure~\ref{LiCl_conductivity_contribution} shows the concentration dependence of the electrophoretic and correlation contributions to the conductivity for aqueous LiCl at $25\celsius$, normalized by the mean-field conductivity. As can be seen, the electrophoretic contribution is significantly larger than the correlation contribution, which is consistent with recent studies that successfully model the conductivity of aqueous electrolytes by focusing primarily on the electrophoretic effect~\cite{vinogradova2025conductivity,vinogradova2023electrophoresis}. Notably, the magnitude of the normalized correlation contribution initially increases, reaches a maximum, and then declines towards zero. This behavior is analogous to that first described by Overbeek \cite{jthg1943theorie} for colloidal systems and recently predicted within the mean-spherical approximation (MSA) framework \cite{bernard2023analytical}.

\FloatBarrier
\subsection{Temperature dependence of conductivity}\label{temp_dep_of_cond}

\begin{figure}[!htbp]
\centering
\subfloat[]{\includegraphics[width=0.46\linewidth]{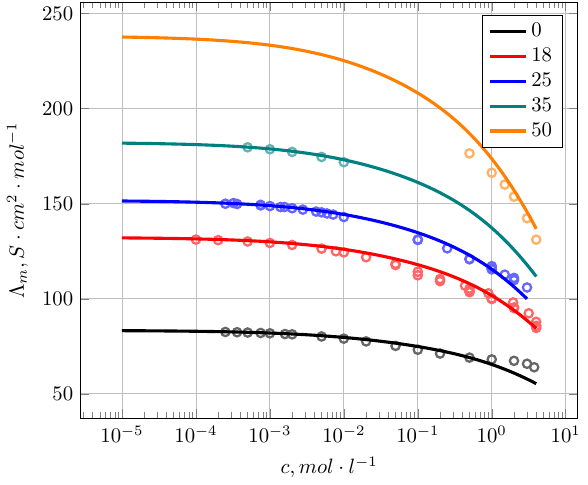}\label{temp_dep_KBr}}\hfill
\subfloat[]{\includegraphics[width=0.46\linewidth]{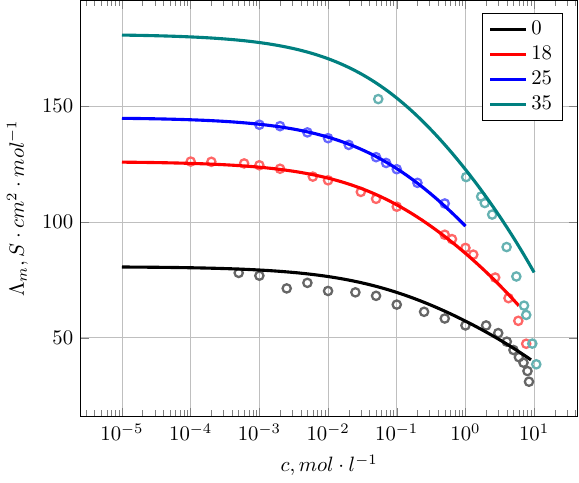}\label{temp_dep_NH4NO3}}\par\smallskip
\subfloat[]{\includegraphics[width=0.46\linewidth]{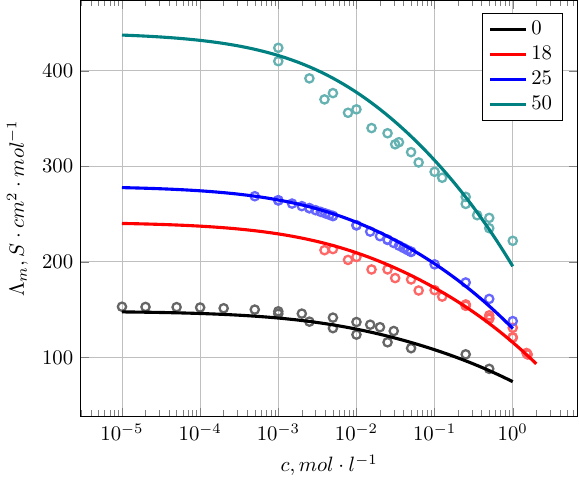}\label{temp_dep_BaCl2}}\hfill
\subfloat[]{\includegraphics[width=0.46\linewidth]{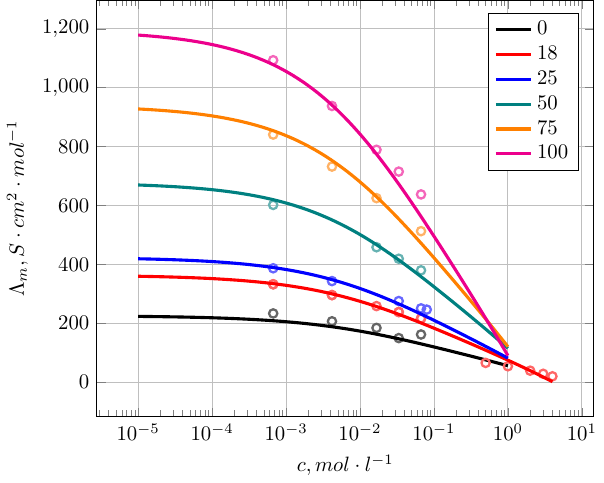}\label{temp_dep_LaNO33}}
\caption{Temperature dependence of molar electrical conductivity as a function of concentration for the following aqueous electrolyte solutions: KBr (a), NH$_4$NO$_3$ (b), BaCl$_2$ (c) and La(NO$_3$)$_3$ (d). Legend entries give temperatures in degrees Celsius.}
\label{temp_dep}
\end{figure}

Let us now investigate the temperature dependence of the conductivity for the case of aqueous electrolyte solutions. Having determined the charge smearing length for each ion from a fit at a specific temperature, we can calculate the conductivity at other temperatures in the range of the solvent-property correlations. The temperature dependence of the pure solvent properties is modeled using established literature relations: dielectric permittivity is described by the equation from Ref.~\cite{malmberg1956dielectric}, and shear viscosity by the relation from Ref.~\cite{laliberte2007model} (see Appendix~\ref{app:water}). The comparison of computed isotherms with experimental ones is presented in Fig.~\ref{temp_dep} for several aqueous electrolyte solutions -- KBr, NH$_4$NO$_3$, BaCl$_2$ and La(NO$_3$)$_3$ -- whose experimental data were also taken from the compilation \cite{Holze2016Electrochemistry}. As shown, the model predictions agree well with the experimental data, verifying the concept of a temperature-independent charge smearing length, at least within the considered temperature range.

\FloatBarrier
\subsection{Non-aqueous electrolyte solutions}
\begin{figure}[!htbp]
\centering
\subfloat[]{\includegraphics[width=0.46\linewidth,height=4.9cm,keepaspectratio]{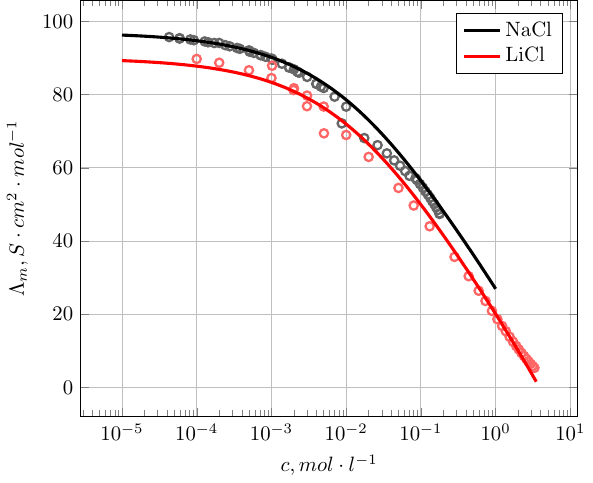}\label{chlorides_in_methanol}}\hfill
\subfloat[]{\includegraphics[width=0.46\linewidth,height=4.9cm,keepaspectratio]{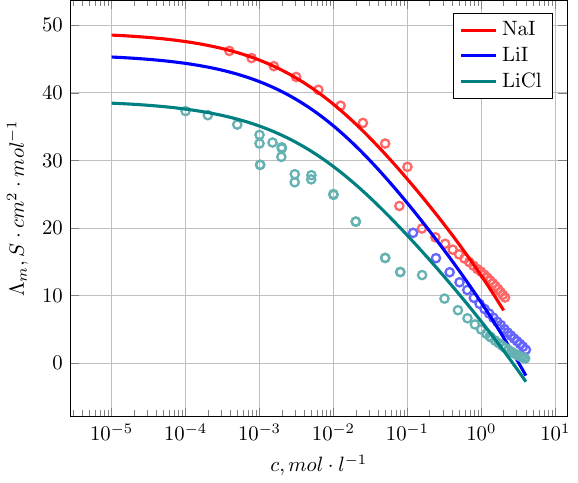}\label{in_ethanol}}\par\smallskip
\subfloat[]{\includegraphics[width=0.46\linewidth,height=4.9cm,keepaspectratio]{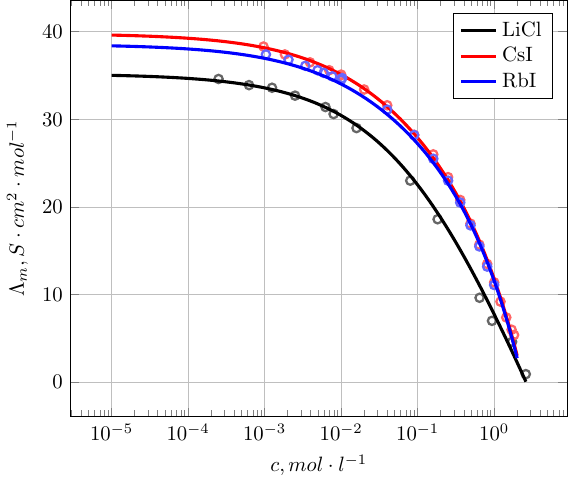}\label{in_DMSO}}\hfill
\subfloat[]{\includegraphics[width=0.46\linewidth,height=4.9cm,keepaspectratio]{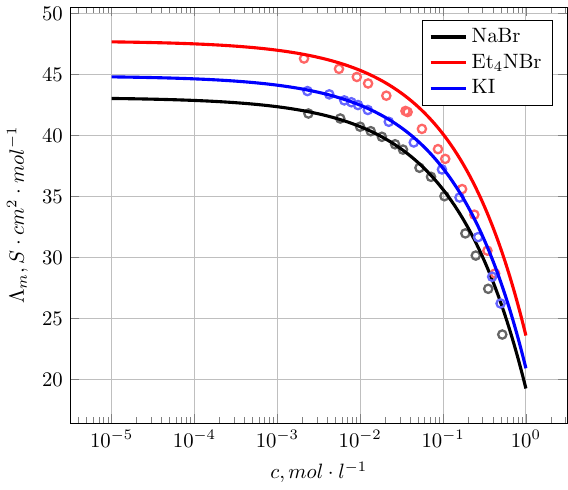}\label{in_n-methylformamide}}\par\smallskip
\subfloat[]{\includegraphics[width=0.46\linewidth,height=4.9cm,keepaspectratio]{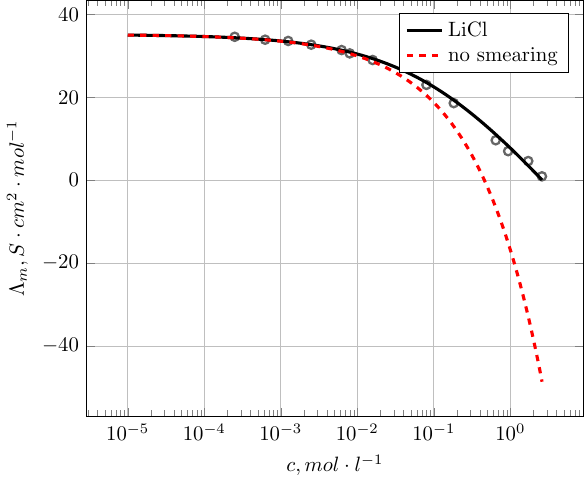}\label{no_smearing_in_DMSO}}\hfill
\subfloat[]{\includegraphics[width=0.46\linewidth,height=4.9cm,keepaspectratio]{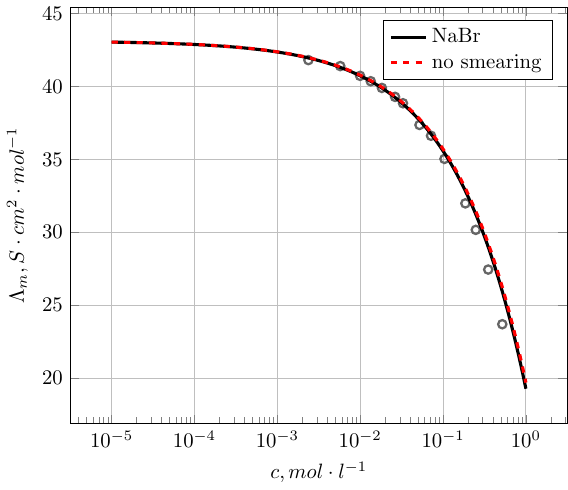}\label{no_smearing_in_n-methylformamide}}
\caption{Molar electrical conductivity as a function of concentration for a set of electrolyte solutions in different solvents: methanol (a), ethanol (b), dimethyl sulfoxide (c) and N-methylformamide (d). Panels (e) and (f) compare the smeared-charge model (solid) and its point-charge limit (dashed) for LiCl in dimethyl sulfoxide and NaBr in N-methylformamide, respectively. Symbols denote measurements.}
\label{non_aqueous_electrolytes}
\end{figure}

It is also instructive to examine the model's transferability to different solvents. Fig.~\ref{non_aqueous_electrolytes} shows the fitted conductivities for a set of electrolyte solutions in several non-aqueous solvents. Experimental conductivity data were taken primarily from the handbook~\cite{janz2012nonaqueous} and Ref.~\cite{dorn2024electrical}, concentration dependence of density was taken from Refs.~\cite{reiser2015density,reiser2016density}. The hydrodynamic radii of the ions were determined from the single-ion conductivities at infinite dilution, as provided in the monograph~\cite{Krestov1987IonicSolvation}. {The resulting fitting parameters for the model in dimethyl sulfoxide, methanol, and ethanol solutions are listed in Table~\ref{nonaqueous_fitting_parameters}. In contrast, for N-methylformamide solutions, fitting parameters converged to an anomalously small value (on the order of $10^{-15}$ m). This outcome can be attributed to two factors. First, the available experimental conductivity data for concentrated solutions in this solvent are limited, confining the fit to a relatively narrow concentration range. Second, N-methylformamide possesses a high permittivity (more than double that of water). This significantly weakens ion-ion correlations, extending the concentration range where the standard DHO theory provides an adequate description. To support this interpretation, we compare model predictions for LiCl in dimethyl sulfoxide (Fig.~\ref{non_aqueous_electrolytes}(e)) and NaBr in N-methylformamide (Fig.~\ref{non_aqueous_electrolytes}(f)), calculated both with and without charge smearing (the latter case corresponds to the point-charge DHO theory). For N-methylformamide, the point-charge theory already describes the data in the measured concentration range. The negative values of the point-charge curve for LiCl in dimethyl sulfoxide occur outside the range of validity of the dilute limiting law. Therefore, obtaining experimental data for more concentrated solutions is essential to constrain the effective charge smearing lengths and establish a more rigorous physical interpretation of the charge smearing length (see the discussion below).}

\begin{table}
    \centering
    \begin{tabular}{c|c|c|c}
        Ion &  \multicolumn{3}{c}{$l_a$, [\angstrom]}        \\
        \hline
                                  &   DMSO &    methanol    & ethanol \\
        \hline
        Li$^+$                    &   0.71 &        1.60    &  2.22   \\
        Na$^+$                    &    -    &        1.64    &  2.46   \\
        K$^+$                     &    -    &        1.37    &   -      \\
        Cs$^+$                    &   1.95 &         -       &   -      \\
        Rb$^+$                    &   2.31 &         -       &   -      \\
        Cl$^{-}$                  &   0.69 &        1.67    &   1.83  \\
        Br$^{-}$                  &   -     &       1.40    &   -  \\
        I$^{-}$                   &   0.04 &        1.70    &   1.35  \\
    \end{tabular}
    \caption{Values of charge smearing length, $l_a$, $[\angstrom]$, for ions, obtained by fitting the conductivity data in different solvents.}
    \label{nonaqueous_fitting_parameters}
\end{table}

\FloatBarrier
\section{Discussion}

The self-consistent Debye--H\"uckel--Onsager (SCDHO) theory explicitly takes into account the non-local nature of ionic charge distributions using a Slater-type charge form factor model. This approach provides a more realistic description of short-range electrostatic interactions in electrolyte solutions, addressing a key limitation of classical point-charge models, especially at high concentrations where the inter-ionic distances are comparable to the spatial extent of the charge clouds. The model successfully reproduces experimental molar conductivity data for a wide range of electrolytes (1:1, 2:1, 3:1, and 1:2), including highly concentrated systems where existing theories often fail. The same ion-specific smearing lengths describe several electrolyte systems in a given solvent (Tables~\ref{fitting_parameters} and \ref{nonaqueous_fitting_parameters}), underscoring the physical robustness of our model.

A notable feature of the SCDHO theory is its ability to decompose total conductivity into mean-field, correlation, and electrophoretic components. As shown in Fig.~\ref{LiCl_conductivity_contribution} for LiCl, the electrophoretic contribution becomes increasingly dominant with concentration, while the correlation term becomes negligible at high concentrations. This trend is physically intuitive, as hydrodynamic drag effects are expected to be more pronounced in concentrated solutions. These findings align with recent studies that either entirely neglect the correlation contribution~\cite{vinogradova2023electrophoresis,vinogradova2025conductivity} or question its significance altogether~\cite{kimura2024revisit}. Furthermore, the minor role of the correlation correction in our results is consistent with predictions from integral equation theory in the mean-spherical approximation (MSA)~\cite{bernard2023analytical}, but contrasts with stochastic density functional theory (SDFT)~\cite{avni2022conductance,avni2022conductivity_}, where the correlation and electrophoretic contributions are predicted to be comparable in magnitude.

The model also captures the temperature dependence of conductivity for different electrolytes (Fig.~\ref{temp_dep}), demonstrating its utility beyond standard ambient conditions. This is achieved by incorporating physically motivated corrections for temperature-dependent solvent permittivity, viscosity, and hydrodynamic radii, without introducing additional fitting parameters.
Furthermore, the extension of the model to non-aqueous electrolyte solutions (Fig.~\ref{non_aqueous_electrolytes}) additionally confirms its robustness. The consistent performance across solvents with different dielectric properties and viscosities indicates that the SCDHO framework is not limited to aqueous systems, but is broadly applicable to a wide range of polar media.
In comparison to earlier modifications of the Debye--H\"uckel--Onsager theory, our approach offers a more fundamental understanding of the underlying charge distribution of ions. By incorporating a charge smearing model within a classical DFT framework, we bridge the gap between microscopic electrostatic structure of ions and macroscopic ionic conductivity.

{The charge smearing length $l_a$ is a fitting parameter whose value inherently depends on the concentration range of the experimental data used. This fact is particularly crucial for non-aqueous solutions where data at high concentrations are scarce. Despite this empirical aspect, it carries significant physical meaning. The fitted values of $l_a$ reflect an interplay of electrostatic and solvation effects, and are not determined by solvent permittivity alone. High permittivity generally enhances screening, reducing the need for a large smearing length -- exemplified by the limiting case of $N$-methylformamide ($l_a \approx 0$). Nonetheless, the ion-specific nature of $l_a$ reveals solvent-dependent solvation physics. For instance, alkali cations exhibit larger $l_a$ in ethanol than in water, attributable to ethanol's lower permittivity and more diffuse solvation layer~\cite{zeng2014solvation}. Conversely, halide anions show larger $l_a$ in water than in methanol, reflecting their more diffuse intrinsic charge distribution and looser aqueous solvation structure~\cite{sarkar2015solvation}, contrasted with the sharper effective charge profile induced by dominant short-range repulsion in methanol.}

{Systematic trends in $l_a$ across ions, most clearly apparent from the extensive data in aqueous solutions, reinforce its chemical interpretability. Monovalent anions like $\mathrm{Br}^{-}$ and $\mathrm{I}^{-}$ possess substantially larger $l_a$ values than smaller anions ($\mathrm{F}^{-}$, $\mathrm{Cl}^{-}$) or alkali cations. This trend is understood through interrelated factors: heavier halides have more diffuse and polarizable electron clouds~\cite{ruffa1963theory}, and in aqueous solution, they form weaker, more disordered hydration shells~\cite{tissandier1998proton,bakker2008structural,marx2010aqueous}. Consequently, $l_a$ increases in the order $\mathrm{F}^{-} < \mathrm{Cl}^{-} < \mathrm{Br}^{-} < \mathrm{I}^{-}$, mirroring established sequences in effective ionic polarizability and solvation free energies. Thus, $l_a$ transcends its role as an adjustable parameter to capture meaningful chemical distinctions between ions.}

{The Slater-type form factor, central to the model, serves as a coarse-grained representation with a dual interpretation. Primarily, it regularizes the Coulomb potential at short distances, avoiding the unphysical divergence of point-charge models and mimicking the spatial delocalization of an ion's charge due to its electronic structure and solvation shell. Secondarily, it phenomenologically incorporates effects beyond pure electrostatics, such as excluded volume of the ion with its solvation shell, specific short-range interactions (e.g., ion pairing), and solvent-structure-induced repulsion. The resulting effective pair potential can thus be viewed as that of a 'dressed' ion~\cite{kjellander1994dressed} in the spirit of coarse-grained descriptions used in soft matter simulations~\cite{warren2014screening,chiacchiera2024coarse}.}

{Moreover, incorporation of the ion-specific smearing lengths $l_a$ in the model justifies the applicability of the RPA, which, while being a linear response theory, may not be adequate for strongly correlated systems, such as multivalent electrolytes. This parameter effectively absorbs essential short-range correlations and non-linear effects (like charge ordering and ion pairing) that are omitted in pure RPA, softening the Coulomb interaction at short distances. The synergy of RPA and $l_a$ yields a computationally efficient model that captures the dominant influence of these correlations on conductivity via a regularized potential, achieving accuracy without the complexity of more rigorous integral-equation theories.}

{Future efforts to refine the model should focus on reducing its empirical component. First-principles calculations could disentangle the contributions bundled into $l_a$, enabling the development of ab initio effective ion-ion interaction potentials. Integrating such potentials -- along with direct correlation functions computed from advanced integral equation theories -- into the SCDHO framework would pave the way for a fully predictive theory of conductivity. Such a theory would be extendable to new ions and solvents without prior experimental fitting.}

A promising and important avenue for future research is the extension of the SCDHO theory to multi-electrolyte solutions. Such systems, containing more than two types of ions, are prevalent in biological systems, geological brines, and industrial processes. The theoretical framework we have developed is general in principle and can be applied beyond the one-component case discussed in Section~\ref{IIB}. However, applying it to multi-electrolyte systems introduces significant complexity due to the combinatorial increase in pairwise correlation functions. Solving such a system would require numerical matrix inversion in the space of ionic species. Despite this challenge, such an extension is computationally feasible and would enable the investigation of phenomena such as the predictive modelling of conductivity in physiological solutions or mixed-salt electrolytes, where specific ion effects and cross-correlations play a critical role. This represents a substantial advance beyond the current state of the art, providing a first-principles tool for engineering complex electrolyte mixtures. The single-component parameters determined in this work will serve as a foundational basis for predicting the electrical conductivity of multi-electrolyte solutions.

\section{Conclusions}
In this study, we have introduced a Self-Consistent Debye--H\"uckel--Onsager theory for the electrical conductivity of electrolyte solutions. The central advancement of this approach is the consistent incorporation of the non-local charge distribution of ions through a Slater-type form factor model. This leads to a regularized Coulomb potential at short ranges, moving beyond the limitations of the classical point-charge approximation.

We have derived a closed set of self-consistent equations for the steady-state ionic correlation functions, starting from the principles of classical Density Functional Theory and the Ornstein--Zernike formalism. As in the original Debye--H\"uckel--Onsager theory, this framework allows us to divide the total conductivity into its three components: mean-field conductivity, correlation conductivity, and electrophoretic conductivity. The developed model, combined with the random phase approximation, yields analytical expressions for the conductivity. It has been successfully validated against a vast set of experimental data for aqueous solutions, accurately reproducing the molar conductivity of 1:1, 2:1, 3:1, and 1:2 electrolytes over a wide range of concentrations. By fitting the model to experimental data, we have obtained a set of ion-specific charge smearing lengths. These parameters, which act as unique identifiers for each ion in a given solvent, are shown to be transferable across different electrolyte systems, highlighting the physical consistency of the model. The theory correctly captures the temperature dependence of conductivity in aqueous solutions. Furthermore, it has been successfully extended to non-aqueous solvents (methanol, ethanol, dimethyl sulfoxide and N-methylformamide), proving its general applicability beyond aqueous systems.

In conclusion, the Self-Consistent Debye--H\"uckel--Onsager theory presented here provides a significant advancement in the molecular-scale modelling of electro-transport in electrolyte solutions. It offers a robust, predictive, and physically grounded tool for calculating conductivity across a wide range of ion valences, concentrations, temperatures, and solvent environments. This work lays the foundation for more precise conductivity prediction in a variety of fields, from electrochemistry and energy storage to biophysics and industrial process design.

\textbf{Acknowledgements.}
Numerical calculations were performed using the computational resources of the HPC facilities at NRU HSE. The article was prepared within the framework of the Mirror Laboratories project at HSE University.

\appendix
\section{Binary electrolyte: pair correlations}\label{Derivation}

\subsection*{\texorpdfstring{Equilibrium relation between $\bar{h}_{12}$ and $c_{12}$}{Equilibrium pair correlations}}

For a binary electrolyte, the Ornstein--Zernike equation in Fourier space reads
\begin{equation}
\bar{h}_{ab}(\mathbf{k}) = c_{ab}(\mathbf{k}) + \sum_{l=1,2} \bar{c}_l \, c_{al}(\mathbf{k}) \, \bar{h}_{lb}(\mathbf{k}), \qquad a,b=1,2 .
\end{equation}
In matrix form, with $\mathbf{h} = (\bar{h}_{ab})$, $\mathbf{c} = (c_{ab})$ and $\bar{\mathbf{C}} = \mathrm{diag}(\bar{c}_1,\bar{c}_2)$, it becomes
\begin{equation}
\mathbf{h} = \mathbf{c} + \mathbf{c}\,\bar{\mathbf{C}}\,\mathbf{h}
\quad\Longrightarrow\quad
\bigl(\mathbf{I} - \mathbf{c}\,\bar{\mathbf{C}}\bigr)\,\mathbf{h} = \mathbf{c}.
\end{equation}
The matrix $\mathbf{M} \equiv \mathbf{I} - \mathbf{c}\,\bar{\mathbf{C}}$ is
\begin{equation}
\mathbf{M}=
\begin{pmatrix}
1-\bar{c}_1c_{11} & -\bar{c}_2c_{12} \\[2pt]
-\bar{c}_1c_{12} & 1-\bar{c}_2c_{22}
\end{pmatrix},
\qquad
\det\mathbf{M} \equiv \Delta =
(1-\bar{c}_1c_{11})(1-\bar{c}_2c_{22})-\bar{c}_1\bar{c}_2c_{12}^2 .
\end{equation}
Inverting $\mathbf{M}$ and computing $\mathbf{h} = \mathbf{M}^{-1}\mathbf{c}$ gives for the off-diagonal elements
\begin{equation}
\bar{h}_{12}=\bar{h}_{21}= \frac{1}{\Delta}\Bigl[(1-\bar{c}_2c_{22})c_{12} + \bar{c}_2c_{12}c_{22}\Bigr]
        = \frac{c_{12}}{\Delta},
\end{equation}
which is Eq.~\eqref{hbar12}.

\subsection*{\texorpdfstring{Field-induced distortion $\omega_{12}$}{Field-induced pair distortion}}
Taking $(a,b)=(1,2)$ in Eq.~\eqref{basic_} we obtain
\begin{equation}
\begin{split}
&(m_1+m_2)k_B Tk^2\omega_{12}(\boldsymbol{k})
- m_1k_B T k^2\bigl[c_{11}(\boldsymbol{k})\bar{c}_1\omega_{12}(\boldsymbol{k})+c_{12}(\boldsymbol{k})\bar{c}_2\omega_{22}(\boldsymbol{k})\bigr] \\
&- m_2k_B T k^2\bigl[c_{21}(\boldsymbol{k})\bar{c}_1\omega_{11}(-\boldsymbol{k})+c_{22}(\boldsymbol{k})\bar{c}_2\omega_{21}(-\boldsymbol{k})\bigr]
= -(q_1m_1-q_2m_2)i(\mathbf{E}\cdot\mathbf{k})\bar{h}_{12}(\boldsymbol{k}),
\end{split}
\end{equation}
With $\omega_{11}=\omega_{22}=0$, $\omega_{ab}(-\boldsymbol{k})=\omega_{ba}(\boldsymbol{k})$ the equation simplifies to
\begin{equation}
\bigl[m_1+m_2-m_1\bar{c}_1c_{11}(\boldsymbol{k})-m_2\bar{c}_2c_{22}(\boldsymbol{k})\bigr]k_B T k^2\omega_{12}(\boldsymbol{k})
= -(q_1m_1-q_2m_2)i(\mathbf{E}\cdot\mathbf{k})\bar{h}_{12}(\boldsymbol{k}).
\end{equation}

Solving for $\omega_{12}$ and inserting $\bar{h}_{12}=c_{12}/\Delta$ from Eq.~\eqref{hbar12} yields
\begin{equation}
\omega_{12}(\mathbf{k}) = -
\frac{(q_1m_1-q_2m_2)\,i(\mathbf{E}\cdot\mathbf{k})\,c_{12}(\mathbf{k})}
{k_B T\,\Delta(\mathbf{k})\,k^{2}\,
\bigl[m_1\bigl(1-\bar{c}_1c_{11}(\mathbf{k})\bigr)+
      m_2\bigl(1-\bar{c}_2c_{22}(\mathbf{k})\bigr)\bigr]}
\end{equation}
which is Eq.~\eqref{omega12}.

\selectlanguage{english}
\FloatBarrier
\section{Supplementary fits and solvent properties}\label{app:supplement}

\subsection{Temperature dependence of water viscosity and permittivity}\label{app:water}
The shear viscosity of pure water was calculated using the correlation of Lalibert\'e~\cite{laliberte2007model}, based on the reference data compiled by Marsh~\cite{marsh1987recommended}:
\begin{equation}
\eta=\frac{t+246}{(0.05594t+5.2842)t+137.37},
\end{equation}
where $\eta$ is in $\mathrm{mPa\,s}$ and $t$ is the temperature in degrees Celsius.

The relative permittivity of pure water was calculated using the fit of Malmberg and Maryott~\cite{malmberg1956dielectric}:
\begin{equation}
\frac{\varepsilon}{\varepsilon_0}=87.740-0.40008t+9.398\times10^{-4}t^2-1.410\times10^{-6}t^3,
\end{equation}
where $\varepsilon_0$ is the vacuum permittivity and $t$ is in degrees Celsius.

\subsection{Aqueous electrolyte solutions}
\begin{figure}[!htbp]
\centering
\subfloat[]{\includegraphics[width=0.46\linewidth]{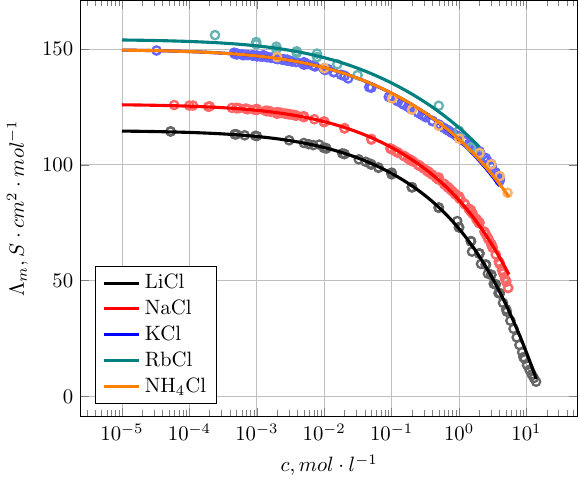}\label{1_1_chlorides}}\hfill
\subfloat[]{\includegraphics[width=0.46\linewidth]{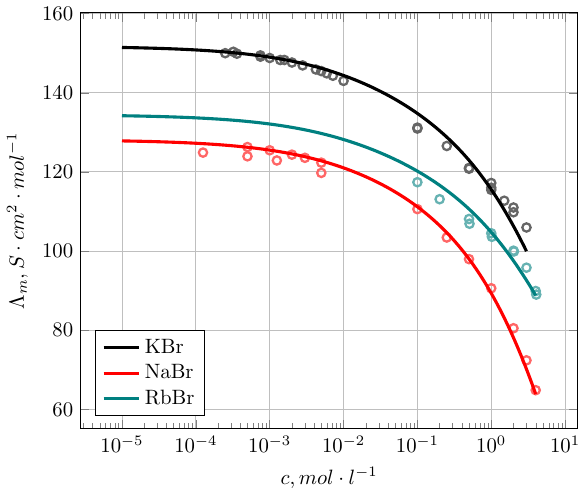}\label{1_1_bromides}}\par\smallskip
\subfloat[]{\includegraphics[width=0.46\linewidth]{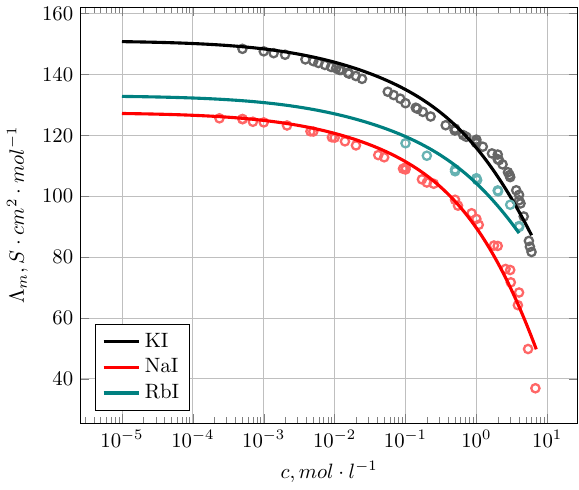}\label{1_1_iodides}}\hfill
\subfloat[]{\includegraphics[width=0.46\linewidth]{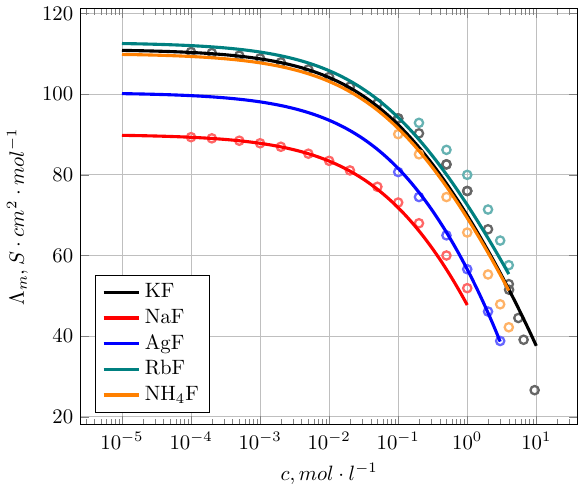}\label{1_1_fluorides}}
\caption{Molar electrical conductivity as a function of concentration for a set of aqueous electrolyte solutions: 1:1 chlorides (a), 1:1 bromides (b), 1:1 iodides (c) and 1:1 fluorides (d).}
\label{1_1_aqueous_electrolytes}
\end{figure}

\begin{figure}[!htbp]
\centering
\subfloat[]{\includegraphics[width=0.46\linewidth]{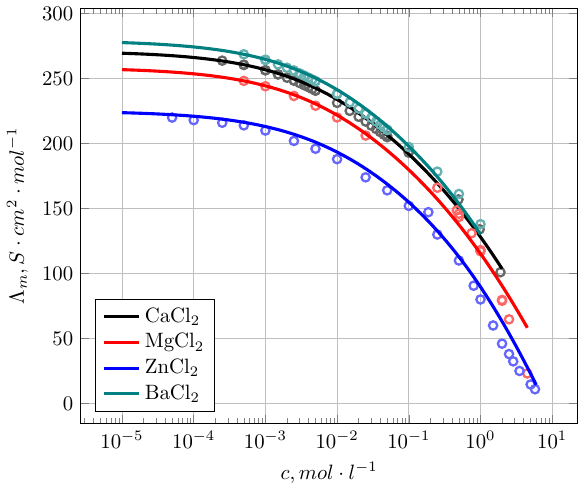}\label{2_1_chlorides}}\hfill
\subfloat[]{\includegraphics[width=0.46\linewidth]{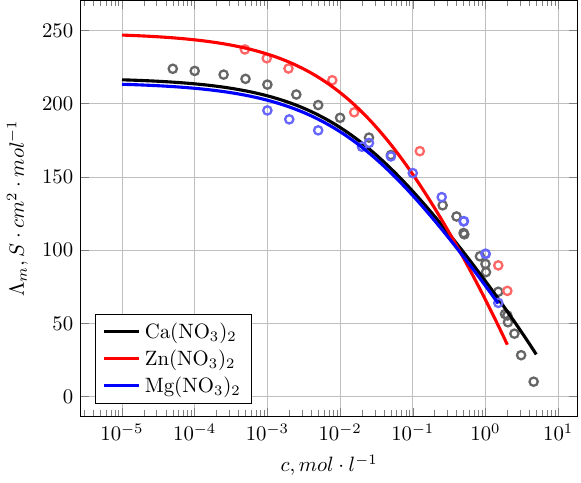}\label{2_1_nitrates}}
\caption{Molar electrical conductivity as a function of concentration for a set of aqueous electrolyte solutions: 2:1 chlorides (a) and 2:1 nitrates (b).}
\label{2_1_aqueous_electrolytes}
\end{figure}

\begin{figure}[!htbp]
\centering
\subfloat[]{\includegraphics[width=0.46\linewidth]{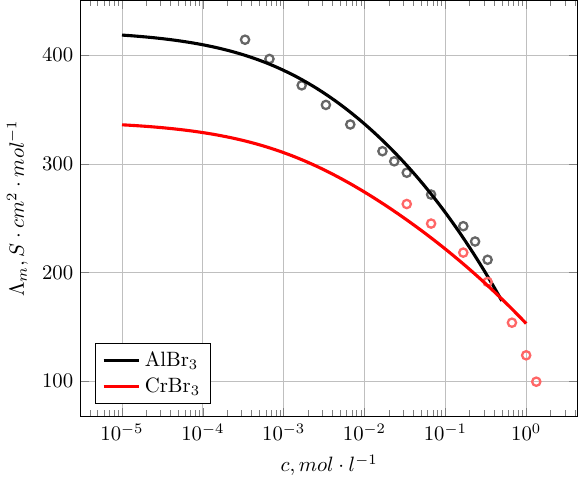}\label{3_1_bromides}}\hfill
\subfloat[]{\includegraphics[width=0.46\linewidth]{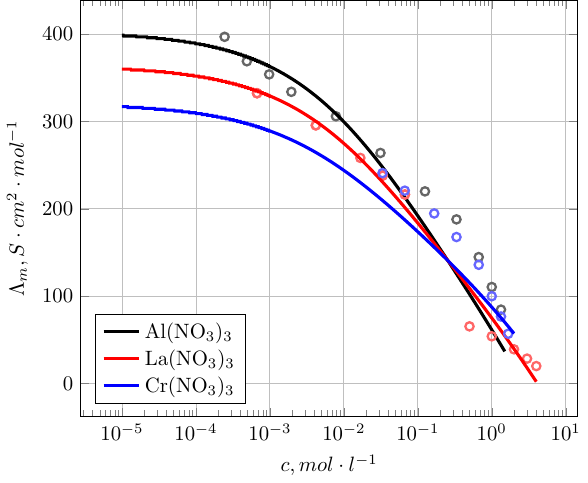}\label{3_1_nitrates}}
\caption{Molar electrical conductivity as a function of concentration for a set of aqueous electrolyte solutions: 3:1 bromides (a) and 3:1 nitrates (b).}
\label{3_1_aqueous_electrolytes}
\end{figure}

\begin{figure}[!htbp]
\centering
\subfloat[]{\includegraphics[width=0.46\linewidth]{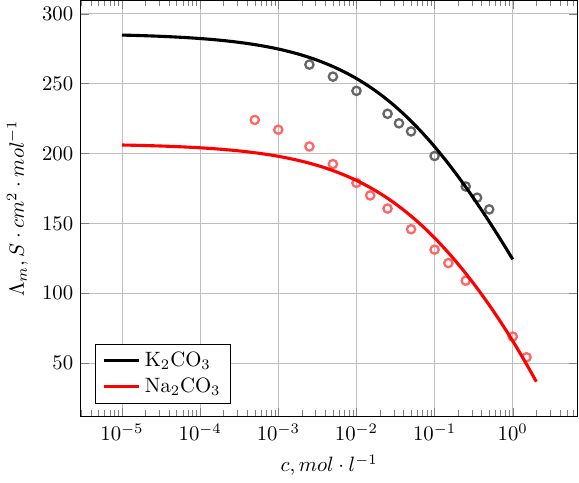}\label{1_2_carbonates}}\hfill
\subfloat[]{\includegraphics[width=0.46\linewidth]{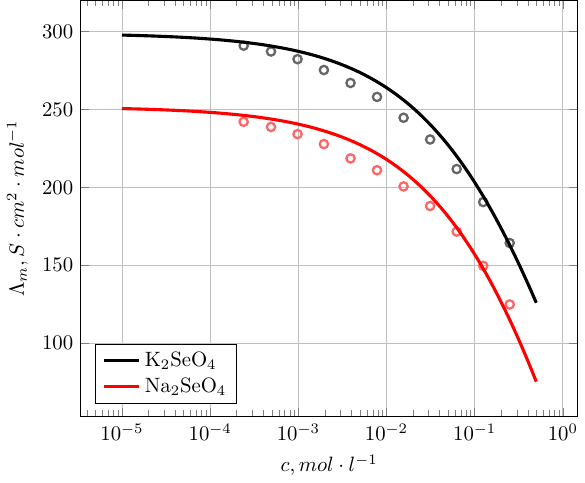}\label{1_2_selenates}}\par\smallskip
\subfloat[]{\includegraphics[width=0.46\linewidth]{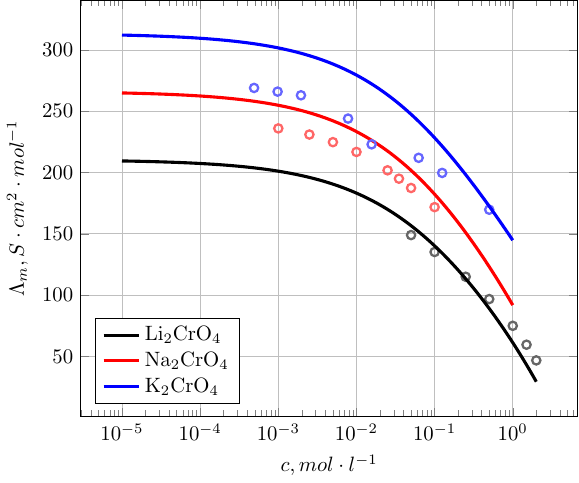}\label{1_2_chromates}}
\caption{Molar electrical conductivity as a function of concentration for a set of aqueous electrolyte solutions: 1:2 carbonates (a), 1:2 selenates (b) and 1:2 chromates (c).}
\label{1_2_aqueous_electrolytes}
\end{figure}

Figures~\ref{1_1_aqueous_electrolytes}--\ref{1_2_aqueous_electrolytes} compare the calculated molar conductivities with the measurements compiled by Holze~\cite{Holze2016Electrochemistry} for 1:1, 2:1, 3:1, and 1:2 aqueous electrolytes. Symbols show measurements and solid lines show the fits. Table~\ref{fitting_parameters} lists the fitted charge smearing lengths.

\FloatBarrier
\subsection{Non-aqueous electrolyte solutions}

\begin{figure}[!htbp]
\centering
\subfloat[]{\includegraphics[width=0.46\linewidth]{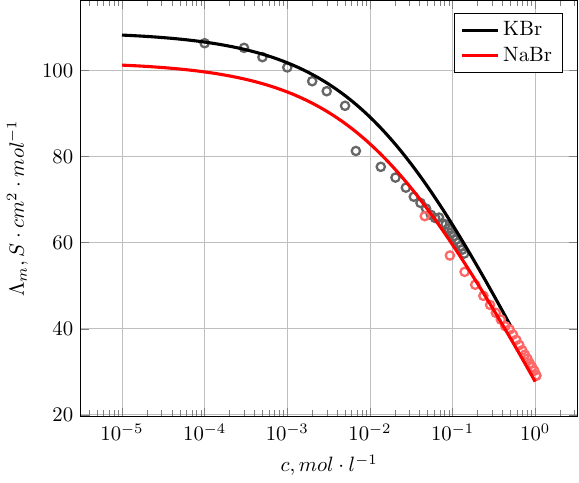}\label{bromides_in_methanol}}\hfill
\subfloat[]{\includegraphics[width=0.46\linewidth]{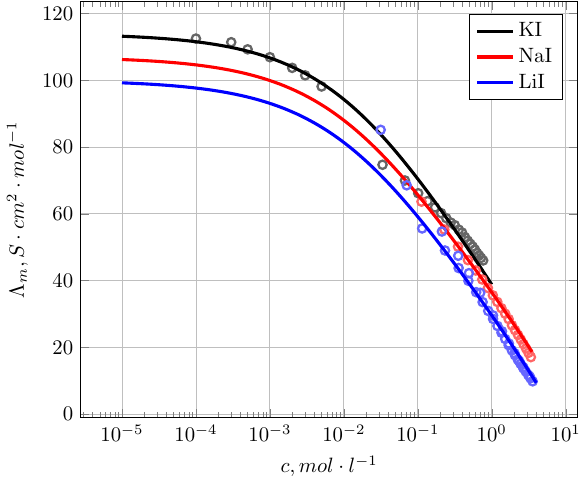}\label{iodides_in_methanol}}
\caption{Molar electrical conductivity as a function of concentration for a set of methanol electrolyte solutions: bromides (a) and iodides (b).}
\label{methanol_supplement}
\end{figure}

\begin{figure}[!htbp]
  \centering
  \includegraphics[width=0.54\linewidth]{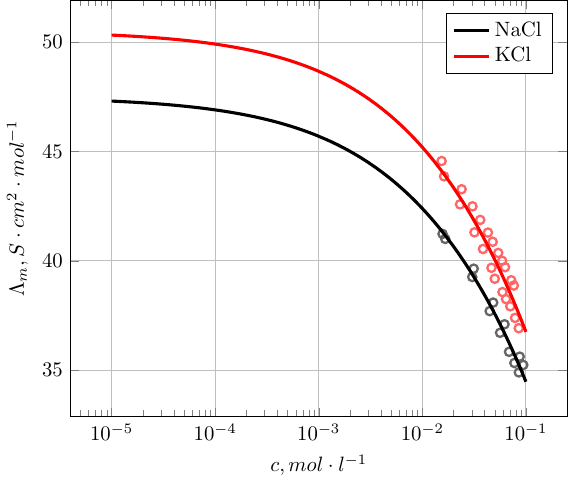}
\caption{Molar electrical conductivity versus salt concentration for NaCl and KCl in formic acid.}
\label{formic_acid_supplement}
\end{figure}

\FloatBarrier
\bibliography{name}

@article{schoen2007mechanism,
  title={The mechanism of extracellular stimulation of nerve cells on an electrolyte-oxide-semiconductor capacitor},
  author={Schoen, Ingmar and Fromherz, Peter},
  journal={Biophysical journal},
  volume={92},
  number={3},
  pages={1096--1111},
  year={2007},
  publisher={Elsevier}
}

@article{bakker2008structural,
  title={Structural dynamics of aqueous salt solutions},
  author={Bakker, HJ},
  journal={Chemical reviews},
  volume={108},
  number={4},
  pages={1456--1473},
  year={2008},
  publisher={ACS Publications}
}

@article{marx2010aqueous,
  title={Aqueous basic solutions: hydroxide solvation, structural diffusion, and comparison to the hydrated proton},
  author={Marx, Dominik and Chandra, Amalendu and Tuckerman, Mark E},
  journal={Chemical reviews},
  volume={110},
  number={4},
  pages={2174--2216},
  year={2010},
  publisher={ACS Publications}
}

@article{ruffa1963theory,
  title={Theory of the electronic polarizabilities of ions in crystals: Application to the alkali halide crystals},
  author={Ruffa, AR},
  journal={Physical Review},
  volume={130},
  number={4},
  pages={1412},
  year={1963},
  publisher={APS}
}

@article{tissandier1998proton,
  title={The Proton's Absolute Aqueous Enthalpy and Gibbs Free Energy of Solvation from Cluster-Ion Solvation Data},
  author={Tissandier, Michael D and Cowen, Kenneth A and Feng, Wan Yong and Gundlach, Ellen and Cohen, Michael H and Earhart, Alan D and Coe, James V and Tuttle Jr, Thomas R},
  journal={J. Phys. Chem. A},
  volume={102},
  pages={7787--7794},
  year={1998}
}

@article{kjellander1994dressed,
  title={Dressed-ion theory for electrolyte solutions: A Debye--H{\"u}ckel-like reformulation of the exact theory for the primitive model},
  author={Kjellander, Roland and Mitchell, D John},
  journal={The Journal of chemical physics},
  volume={101},
  number={1},
  pages={603--626},
  year={1994},
  publisher={American Institute of Physics}
}

@article{schmidt2022power,
  title={Power functional theory for many-body dynamics},
  author={Schmidt, Matthias},
  journal={Reviews of Modern Physics},
  volume={94},
  number={1},
  pages={015007},
  year={2022},
  publisher={APS}
}

@article{robin2024correlation,
  title={Correlation-induced viscous dissipation in concentrated electrolytes},
  author={Robin, Paul},
  journal={The Journal of Chemical Physics},
  volume={160},
  number={6},
  year={2024},
  publisher={AIP Publishing}
}

@book{Lechner2008Static,
  editor = {M. D. Lechner},
  title = {Static Dielectric Constants of Pure Liquids and Binary Liquid Mixtures},
  subtitle = {Supplement to IV/6},
  series = {Landolt-B\"ornstein: Numerical Data and Functional Relationships in Science and Technology - New Series},
  publisher = {Springer Berlin, Heidelberg},
  year = {2008},
  edition = {1},
  isbn = {978-3-540-75505-0},
  doi = {10.1007/978-3-540-75506-7},
  copyright = {Springer-Verlag Berlin Heidelberg 2008},
  topics = {Physics, general, Classical and Continuum Physics, Physical Chemistry}
}

@book{Wohlfarth2016Viscosity,
  author = {Christian Wohlfarth},
  editor = {M. D. Lechner},
  title = {Viscosity of Pure Organic Liquids and Binary Liquid Mixtures},
  series = {Landolt-B\"ornstein: Numerical Data and Functional Relationships in Science and Technology - New Series},
  publisher = {Springer Berlin, Heidelberg},
  year = {2016},
  edition = {1},
  isbn = {978-3-662-49216-1},
  doi = {10.1007/978-3-662-49218-5},
  copyright = {Springer-Verlag Berlin Heidelberg 2017}
}

@article{vinogradova2025conductivity,
  title={Conductivity of concentrated salt solutions},
  author={Vinogradova, Olga I and Silkina, Elena F},
  journal={The Journal of Chemical Physics},
  volume={163},
  number={4},
  year={2025},
  publisher={AIP Publishing}
}

@book{metheny2012fluid,
  title={Fluid and electrolyte balance},
  author={Metheny, Norma},
  year={2012},
  publisher={Jones \& Bartlett Publishers}
}

@article{reiser2016density,
  title={Density of ethanolic alkali halide salt solutions by experiment and molecular simulation},
  author={Reiser, Steffen and Horsch, Martin and Hasse, Hans},
  journal={Fluid Phase Equilibria},
  volume={408},
  pages={141--150},
  year={2016},
  publisher={Elsevier}
}

@article{sarkar2015solvation,
  title={Solvation structures of lithium halides in methanol--water mixtures},
  author={Sarkar, Atanu and Dixit, Mayank Kumar and Tembe, BL},
  journal={Chemical Physics},
  volume={447},
  pages={76--85},
  year={2015},
  publisher={Elsevier}
}

@article{zeng2014solvation,
  title={Solvation structure and dynamics of Li+ ion in liquid water, methanol and ethanol: A comparison study},
  author={Zeng, Yongping and Wang, Chunfeng and Zhang, Xiaobin and Ju, Shengui},
  journal={Chemical Physics},
  volume={433},
  pages={89--97},
  year={2014},
  publisher={Elsevier}
}

@book{marsh1987recommended,
  title={Recommended Reference Materials for the Realization of Physicochemical Properties},
  editor={Marsh, Kenneth N.},
  year={1987},
  publisher={Blackwell Scientific Publications Oxford}
}

@article{reiser2015density,
  title={Density of methanolic alkali halide salt solutions by experiment and molecular simulation},
  author={Reiser, Steffen and Horsch, Martin and Hasse, Hans},
  journal={Journal of Chemical \& Engineering Data},
  volume={60},
  number={6},
  pages={1614--1628},
  year={2015},
  publisher={ACS Publications}
}

@article{dorn2024electrical,
  title={Electrical conductivity of lithium, sodium, potassium, and quaternary ammonium salts in water, acetonitrile, methanol, and ethanol over a wide concentration range},
  author={Dorn, Marvin and Kareth, Sabine and Weidner, Eckhard and Petermann, Marcus},
  journal={Journal of Chemical \& Engineering Data},
  volume={69},
  number={4},
  pages={1493--1502},
  year={2024},
  publisher={ACS Publications}
}

@article{bohr1964electrolytes,
  title={Electrolytes and smooth muscle contraction},
  author={BOHR, DAVID F},
  journal={Pharmacological reviews},
  volume={16},
  number={1},
  pages={85--111},
  year={1964},
  publisher={ASPET}
}

@article{onsager2002irreversible,
  title={Irreversible processes in electrolytes. Diffusion, conductance and viscous flow in arbitrary mixtures of strong electrolytes},
  author={Onsager, Lars and Fuoss, Raymond Matthew},
  journal={The Journal of Physical Chemistry},
  volume={36},
  number={11},
  pages={2689--2778},
  year={1932},
  publisher={ACS Publications}
}

@book{Holze2016Electrochemistry,
  author = {Rudolf Holze},
  editor = {M. D. Lechner},
  title = {Electrochemistry},
  subtitle = {Subvolume B: Electrical Conductivities and Equilibria of Electrochemical Systems - Part 2: Deep Eutectic Solvents and Electrolyte Solutions},
  series = {Landolt-B\"ornstein: Numerical Data and Functional Relationships in Science and Technology - New Series},
  publisher = {Springer Berlin, Heidelberg},
  year = {2016},
  edition = {1},
  isbn = {978-3-662-49249-9},
  doi = {10.1007/978-3-662-49251-2},
  copyright = {Springer-Verlag Berlin Heidelberg 2016},
  topics = {Electrochemistry, Phase Transitions and Multiphase Systems, Surface and Interface Science, Thin Films, Energy Systems}
}

@book{janz2012nonaqueous,
  title={Nonaqueous electrolytes handbook},
  author={Janz, George J},
  volume={1},
  year={2012},
  publisher={Elsevier}
}

@book{Krestov1987IonicSolvation,
  author = {G. A. Krestov and N. P. Novoselov and I. S. Perelygin and A. M. Kolker and L. P. Safonova and V. D. Ovchinnikova and V. N. Trostin},
  title = {Ionic Solvation},
  series = {Problems of Solution Chemistry Series},
  publisher = {G. A. Krestov Institute of Solution Chemistry of the Russian Academy of Sciences},
  year = {1987},
  address = {Moscow},
  language = {russian},
  note = {In Russian},
}

@article{mao2015modeling,
  title={Modeling aggregation of ionic surfactants using a smeared charge approximation in dissipative particle dynamics simulations},
  author={Mao, Runfang and Lee, Ming-Tsung and Vishnyakov, Aleksey and Neimark, Alexander V},
  journal={The Journal of Physical Chemistry B},
  volume={119},
  number={35},
  pages={11673--11683},
  year={2015},
  publisher={ACS Publications}
}

@book{budkov2024statistical,
  title={Statistical Field Theory of Ion-Molecular Fluids: Fundamentals and Applications in Physical Chemistry and Electrochemistry},
  author={Budkov, Yury A and Kalikin, Nikolai N},
  year={2024},
  publisher={Springer Nature}
}

@article{adar2019screening,
  title={Screening length for finite-size ions in concentrated electrolytes},
  author={Adar, Ram M and Safran, Samuel A and Diamant, Haim and Andelman, David},
  journal={Physical Review E},
  volume={100},
  number={4},
  pages={042615},
  year={2019},
  publisher={APS}
}

@article{frydel2016double,
  title={The double-layer structure of overscreened surfaces by smeared-out ions},
  author={Frydel, Derek},
  journal={The Journal of chemical physics},
  volume={145},
  number={18},
  year={2016},
  publisher={AIP Publishing}
}

@article{frydel2013double,
  title={The double-layer of penetrable ions: An alternative route to charge reversal},
  author={Frydel, Derek and Levin, Yan},
  journal={The Journal of chemical physics},
  volume={138},
  number={17},
  year={2013},
  publisher={AIP Publishing}
}

@book{barrat2003basic,
  title={Basic concepts for simple and complex liquids},
  author={Barrat, Jean-Louis and Hansen, Jean-Pierre},
  year={2003},
  publisher={Cambridge University Press}
}

@article{fraenkel2018improved,
  title={An improved theory of the electric conductance of ionic solutions based on the concept of the ion-atmosphere's smaller-ion shell},
  author={Fraenkel, Dan},
  journal={Physical Chemistry Chemical Physics},
  volume={20},
  number={47},
  pages={29896--29909},
  year={2018},
  publisher={Royal Society of Chemistry}
}

@article{jthg1943theorie,
  title={Theorie der Elektrophorese},
  author={Overbeek, JTG},
  journal={Kolloid-Beihefte},
  volume={54},
  pages={287},
  year={1943}
}

@article{fraenkel2010simplified,
  title={Simplified electrostatic model for the thermodynamic excess potentials of binary strong electrolyte solutions with size-dissimilar ions},
  author={Fraenkel, Dan},
  journal={Molecular Physics},
  volume={108},
  number={11},
  pages={1435--1466},
  year={2010},
  publisher={Taylor \& Francis}
}

@article{kornyshev1996shape,
  title={The shape of the nonlocal dielectric function of polar liquids and the implications for thermodynamic properties of electrolytes: A comparative study},
  author={Kornyshev, Alexei A and Sutmann, Godehard},
  journal={The Journal of chemical physics},
  volume={104},
  number={4},
  pages={1524--1544},
  year={1996},
  publisher={AIP Publishing}
}

@article{bernard1992conductance,
  title={Conductance in electrolyte solutions using the mean spherical approximation},
  author={Bernard, Olivier and Kunz, Werner and Turq, Pierre and Blum, Lesser},
  journal={The Journal of Physical Chemistry},
  volume={96},
  number={9},
  pages={3833--3840},
  year={1992},
  publisher={ACS Publications}
}

@article{kimura2024revisit,
  title={Revisit the classical theories of the conductance of unassociated electrolytes. Is there reality in relaxation?},
  author={Kimura, Keisaku},
  journal={The Journal of Chemical Physics},
  volume={160},
  number={22},
  year={2024},
  publisher={AIP Publishing}
}

@article{nightingale1959phenomenological,
  title={Phenomenological theory of ion solvation. Effective radii of hydrated ions},
  author={Nightingale Jr, ER},
  journal={The Journal of Physical Chemistry},
  volume={63},
  number={9},
  pages={1381--1387},
  year={1959},
  publisher={ACS Publications}
}

@article{malmberg1956dielectric,
  title={Dielectric Constant of Water from {$0^\circ$} to {$100^\circ$} C},
  author={Malmberg, C. G. and Maryott, A. A.},
  journal={Journal of research of the National Bureau of Standards},
  volume={56},
  number={1},
  pages={1--8},
  year={1956}
}

@article{laliberte2007model,
  title={Model for calculating the viscosity of aqueous solutions},
  author={Lalibert{\'e}, Marc},
  journal={Journal of Chemical \& Engineering Data},
  volume={52},
  number={2},
  pages={321--335},
  year={2007},
  doi={10.1021/je0604075},
  publisher={ACS Publications}
}

@article{warren2014screening,
  title={Screening properties of four mesoscale smoothed charge models, with application to dissipative particle dynamics},
  author={Warren, Patrick B and Vlasov, Andrey},
  journal={The Journal of chemical physics},
  volume={140},
  number={8},
  year={2014},
  publisher={AIP Publishing}
}

@article{kalikin2024modified,
  title={Modified Debye--H{\"u}ckel--Onsager theory for electrical conductivity in aqueous electrolyte solutions: Account of ionic charge nonlocality},
  author={Kalikin, Nikolai N and Budkov, Yury A},
  journal={The Journal of Chemical Physics},
  volume={161},
  number={17},
  year={2024},
  publisher={AIP Publishing}
}

@article{chiacchiera2024coarse,
  title={Coarse-grained polarizable soft solvent models, with applications in dissipative particle dynamics},
  author={Chiacchiera, Silvia and Warren, Patrick B and Masters, Andrew J and Seaton, Michael A},
  journal={The Journal of chemical physics},
  volume={161},
  number={17},
  year={2024},
  publisher={AIP Publishing}
}

@article{ibergay2009electrostatic,
  title={Electrostatic interactions in dissipative particle dynamics: toward a mesoscale modeling of the polyelectrolyte brushes},
  author={Ibergay, Cyrille and Malfreyt, Patrice and Tildesley, Dominic J},
  journal={Journal of Chemical Theory and Computation},
  volume={5},
  number={12},
  pages={3245--3259},
  year={2009},
  publisher={ACS Publications}
}

@article{rubashkin2015nonlocal,
  title={Nonlocal electrostatic theory of ion solvation: A combination of the overscreening effect in the dielectric response of the medium with a smeared ion charge distribution},
  author={Rubashkin, A. A. and Vorotyntsev, M. A. and Antipov, E. M. and Aldoshin, S. M.},
  journal={Doklady Physical Chemistry},
  volume={464},
  pages={198--201},
  year={2015},
  doi={10.1134/S0012501615090031}
}

@book{pitaevskii2012physical,
  title={Physical Kinetics: Volume 10},
  author={Pitaevskii, Lev P and Lifshitz, EM},
  volume={10},
  year={2012},
  publisher={Butterworth-Heinemann}
}

@article{avni2022conductance,
  title={Conductance of concentrated electrolytes: Multivalency and the Wien effect},
  author={Avni, Yael and Andelman, David and Orland, Henri},
  journal={The Journal of Chemical Physics},
  volume={157},
  number={15},
  year={2022},
  publisher={AIP Publishing}
}

@article{onsager1927theory,
  title={On the theory of electrolytes. II},
  author={Onsager, Lars},
  journal={Physikalische Zeitschrift},
  volume={28},
  pages={277--298},
  year={1927}
}

@article{bernard2023analytical,
  title={On analytical theories for conductivity and self-diffusion in concentrated electrolytes},
  author={Bernard, Olivier and Jardat, Marie and Rotenberg, Benjamin and Illien, Pierre},
  journal={The Journal of Chemical Physics},
  volume={159},
  number={16},
  year={2023},
  publisher={AIP Publishing}
}

@article{vinogradova2023electrophoresis,
  title={Electrophoresis of ions and electrolyte conductivity: From bulk to nanochannels},
  author={Vinogradova, Olga I and Silkina, Elena F},
  journal={The Journal of Chemical Physics},
  volume={159},
  number={17},
  year={2023},
  publisher={AIP Publishing}
}

@article{avni2022conductivity_,
  title={Conductivity of concentrated electrolytes},
  author={Avni, Yael and Adar, Ram M and Andelman, David and Orland, Henri},
  journal={Physical Review Letters},
  volume={128},
  number={9},
  pages={098002},
  year={2022},
  publisher={APS}
}

@article{naseri2023new,
  title={New Electrical Conductivity Model for Electrolyte Solutions Based on the Debye-Huckel-Onsager Theory},
  author={Naseri Boroujeni, Saman and Maribo-Mogensen, Bjorn and Liang, Xiaodong and Kontogeorgis, Georgios M},
  journal={The Journal of Physical Chemistry B},
  volume={127},
  number={46},
  pages={9954--9975},
  year={2023},
  publisher={ACS Publications}
}

@article{berthoumieux2024nonlinear,
  title={Nonlinear conductivity of aqueous electrolytes: Beyond the first Wien effect},
  author={Berthoumieux, H{\'e}l{\`e}ne and D{\'e}mery, Vincent and Maggs, Anthony C},
  journal={The Journal of Chemical Physics},
  volume={161},
  number={18},
  year={2024},
  publisher={AIP Publishing}
}

@article{budkov2025fluctuation,
  title={A Fluctuation Theory of Liquid-Phase Solutions: Shear Viscosity},
  author={Budkov, Yury A and Kalikin, Nikolai N and Brandyshev, Petr E},
  journal={The Journal of Chemical Physics},
  volume={163},
  number={2},
  year={2025},
  publisher={AIP Publishing}
}

@article{illien2025dean,
  title={The Dean--Kawasaki equation and stochastic density functional theory},
  author={Illien, Pierre},
  journal={Reports on Progress in Physics},
  volume={88},
  number={8},
  pages={086601},
  year={2025},
  publisher={IOP Publishing}
}

@article{barthel1968conductance,
  title={Conductance of electrolyte solutions},
  author={Barthel, J},
  journal={Angewandte Chemie International Edition in English},
  volume={7},
  number={4},
  pages={260--277},
  year={1968},
  publisher={Wiley Online Library}
}

@article{debye1923theory,
  title={Theory of electrolytes-part II: law of the limit of electrolytic conduction},
  author={Debye, P and H{\"u}ckel, E},
  journal={Phys. Z},
  volume={24},
  pages={305},
  year={1923}
}

@article{alkhadra2022electrochemical,
  title={Electrochemical methods for water purification, ion separations, and energy conversion},
  author={Alkhadra, Mohammad A and Su, Xiao and Suss, Matthew E and Tian, Huanhuan and Guyes, Eric N and Shocron, Amit N and Conforti, Kameron M and De Souza, J Pedro and Kim, Nayeong and Tedesco, Michele and others},
  journal={Chemical reviews},
  volume={122},
  number={16},
  pages={13547--13635},
  year={2022},
  publisher={ACS Publications}
}

@article{chaplin2019prospect,
  title={The prospect of electrochemical technologies advancing worldwide water treatment},
  author={Chaplin, Brian P},
  journal={Accounts of chemical research},
  volume={52},
  number={3},
  pages={596--604},
  year={2019},
  publisher={ACS Publications}
}

@article{boroujeni2023estimation,
  title={On the estimation of equivalent conductivity of electrolyte solutions: The effect of relative static permittivity and viscosity},
  author={Boroujeni, Saman Naseri and Maribo-Mogensen, Bj{\o}rn and Liang, Xiaodong and Kontogeorgis, Georgios M},
  journal={Fluid Phase Equilibria},
  volume={567},
  pages={113698},
  year={2023},
  publisher={Elsevier}
}

@article{wang2010fluctuation,
  title={Fluctuation in electrolyte solutions: The self energy},
  author={Wang, Zhen-Gang},
  journal={Physical Review E-Statistical, Nonlinear, and Soft Matter Physics},
  volume={81},
  number={2},
  pages={021501},
  year={2010},
  publisher={APS}
}

@article{budkov2020statistical,
  title={Statistical field theory of ion--molecular solutions},
  author={Budkov, Yury A},
  journal={Physical Chemistry Chemical Physics},
  volume={22},
  number={26},
  pages={14756--14772},
  year={2020},
  publisher={Royal Society of Chemistry}
}

@book{kraeft1986quantum,
  title={Quantum statistics of charged particle systems},
  author={Kraeft, Wolf-Dietrich and Kremp, Dietrich and Ebeling, Werner and R{\"o}pke, Gerd},
  year={1986},
  publisher={Springer}
}

@article{onsager1927report,
  title={Report on a revision of the conductivity theory},
  author={Onsager, Lars},
  journal={Transactions of the Faraday Society},
  volume={23},
  pages={341--349},
  year={1927},
  publisher={Royal Society of Chemistry}
}

@article{demery2016conductivity,
  title={The conductivity of strong electrolytes from stochastic density functional theory},
  author={D{\'e}mery, Vincent and Dean, David S},
  journal={Journal of Statistical Mechanics: Theory and Experiment},
  volume={2016},
  number={2},
  pages={023106},
  year={2016},
  publisher={IOP Publishing}
}

@article{fuoss1978review,
  title={Review of the theory of electrolytic conductance},
  author={Fuoss, Raymond M},
  journal={Journal of Solution Chemistry},
  volume={7},
  pages={771--782},
  year={1978},
  publisher={Springer}
}

\end{document}